\pdfoutput=1
\documentclass[useAMS,usenatbib]{mn2e}
\usepackage{graphicx} 
\usepackage{dcolumn}  
\usepackage{bm}       
\usepackage{amssymb,amsmath}  
\usepackage{color}

\title[RSD and BAO with photometric LRGs from SDSS]
{Clustering of photometric luminous red galaxies I :
Growth of Structure and Baryon Acoustic Feature}

\author[M. Crocce et
al.]{M. Crocce$^1$\thanks{E-mail:martincrocce@gmail.com},
  E. Gazta\~naga$^1$, A. Cabr\'e$^2$, A. Carnero$^3$, E. S\'anchez$^3$\\
 $^1$Institut de Ci\`encies de l'Espai (IEEC-CSIC), Barcelona, Spain \\
 $^2$University of Pennsylvania, Philadelphia, USA\\
 $^3$Centro de Investigaciones En\'ergeticas, Medioambientales y Tecnol\'ogicas (CIEMAT), Madrid, Spain\\
}

\begin{document}

\date{\today}
\pagerange{1--10} \pubyear{2011}
\maketitle

\begin{abstract}
The possibility of measuring redshift space (RSD) distortions using photometric data have been recently 
highlighted. This effect complements and significantly alters the detectability of
baryon acoustic oscillations (BAO) in photometric surveys. 
In this paper
we present measurements of the angular correlation function of luminous red galaxies (LRGs) 
in the photometric catalog of the final data release (DR7) of the
Sloan Digital Sky Survey II (SDSS).
The sample compromise $\sim 1.5\times 10^6$ LRGs distributed in $0.45
< z < 0.65$, with a characteristic photometric error of $\sim 0.05$. 
Our measured correlation centered at $z=0.55$ is in very good agreement with
predictions from standard $\Lambda$CDM in a broad range of angular scales,
$0.5^\circ < \theta < 6^\circ$. We find that the growth of
structure can indeed be robustly measured, with errors matching
expectations. The velocity
growth rate is recovered as $f \sigma_8 = 0.53 \pm
0.42$ when no prior is imposed on the growth factor and the background
geometry follows a $\Lambda$CDM model with WMAP7+SNIa priors.
This is compatible with the corresponding General Relativity (GR) prediction 
$f \sigma_8 = 0.45$ for our fiducial cosmology.
If we adopt a parameterization such that $f=\Omega ^\gamma_m(z)$, with
$\gamma \approx 0.55$ in GR, and 
combine our $f\sigma_8$ measurement with the corresponding ones from
spectroscopic LRGs at lower redshifts we obtain $\gamma=0.54 \pm 0.17$.
In addition we find evidence for the presence of the baryon acoustic feature matching the
amplitude, location and shape of $\Lambda$CDM predictions. 
The photometric BAO feature is detected with $98\%$ confidence level at $z=0.55$.
\end{abstract}

\begin{keywords}
data analysis -- cosmological parameters -- dark energy -- large-scale structure of the universe
\end{keywords}
\section{Introduction}
\label{sec:intro}
The discovery of an accelerated cosmic expansion has become one of 
the biggest puzzles in modern cosmology over the
last 10 years. 
Several scientific probes have been
proposed to understand the nature of this acceleration. 
From ``geometrical'' tests based on measurements of the
distance-redshift relation such as baryon acoustic oscillations (BAO)
or Type Ia supernovae, to ``growth'' tests sensitive to the growth rate of
perturbations such as redshift space distortions (RSD), weak lensing or cluster
abundance.
The success of these probes relies in the implementation of 
massive, and many times dedicated, observational campaigns that will scan a
good fraction of the observable Universe.
Some such surveys will base their science in galaxy redshifts
derived spectroscopically, what provides accurate radial 
positions. Others will instead measure redshift
photometrically. This yields poorer determination of radial positions but
allows to go deeper in redshift and have higher sampling rate. The
later group 
involves the Dark Energy Survey\footnote{\tt www.darkenergysurvey.org} (DES),  the Physics of the
Accelerating Universe collaboration\footnote{\tt www.pausurvey.org} (PAU) and 
the Panoramic Survey Telescope and Rapid Response System\footnote{\tt
  pan-stars.ifa.hawaii.edu} (PanStarrs) as well as
proposals such as the Large Synoptic Survey
Telescope\footnote{\tt www.lsst.org} (LSST) and the imaging component of
ESA/Euclid \footnote{\tt www.euclid-imaging.net} survey.

Perhaps the most exciting results related to the large scale
structure of the Universe to date have been obtained using
spectroscopic data from surveys such as the two
degree field galaxy redshift survey (2dGRS, \cite{2001MNRAS.328.1039C}) the Sloan Digital Sky
Survey (SDSS, \cite{2000AJ....120.1579Y}) or the recent WiggleZ Dark
Energy survey
(\cite{2010MNRAS.401.1429D}). 
This is particularly so in regards to redshift space
distortions and baryon acoustic oscillations, the two topics of this paper.

Redshift space distortions arise
because the receding speed of galaxies with respect to us is due
not only to the Hubble expansion but also to their peculiar velocity.
Hence, galaxy positions inferred with the
Hubble law are modified with respect to their true positions
depending on the local velocity field.
At large scales the net effect results from the relative strength of the intrinsic
clustering (the bias) and the amplitude of velocity
flows set by the conservation of mass through the growth rate of structure parameter $f=\partial \ln D
/\partial \ln a$ (where $D$ is the linear growth factor and $a$ the
cosmological scale factor). Photometric redshift errors are generally assumed to
wash out these distortions. As recently discussed by
\cite{2010MNRAS.407..520N} and \cite{2011MNRAS.tmp..385C} this is
true to a good extent but does not remove the signal completely if one 
splits the data in redshift bins. Thus, RSD from photometric data
could be a sensitive test for the growth of structure as a function
of redshift (\cite{2011arXiv1102.0968R}). This can then be used to
discriminate modifications of Einstein's gravity from dark energy models.

In turn, BAO originate in the tight coupling
between baryons and photons prior to recombination. At the time of
decoupling their ``last scattering'' imprints a well defined comoving scale
in the spatial distribution of baryons and matter of $\sim 100 \,h^{-1}{\rm Mpc}$,
characterized by a slight excess of pairs over random. This scale is
today imprinted in the distribution of galaxies. Again, poor
distance determination because of photometric redshift estimates wash
out this excess, at least in the radial direction. Photometric surveys
should still be capable of detecting this signature in the angular
distribution of galaxies (e.g. \cite{2005MNRAS.363.1329B}).

In this paper we use the {\it angular correlation function} of luminous red
galaxies (LRGs) in the imaging catalog of the final data release (DR7) of SDSS
to address whether growth of structure can be robustly measured with
photometric data despite several sources of systematic errors, low
resolution photo-z and other unknowns. In parallel, we investigate
whether BAO can be observed in the
clustering pattern of LRGs in concordance with model predictions
affected by redshift space distortions. In a companion paper, \cite{bao_paper},
we discuss the cosmological implications of the presence of BAO in 
the clustering signal.

These tests, that extend previous work using angular power spectrum
(\cite{2007MNRAS.378..852P,2007MNRAS.374.1527B,2011MNRAS.412.1669T}), may serve as a proof-of-concept for the potential of future, more
accurate, photometric data to place interesting constraints into the 
nature of cosmic expansion, and/or provide valuable higher redshift 
leverage to complement spectroscopic measurements. 

This paper is organized as follows. In Sec.~\ref{sec:sel} we discuss
our data, including the selection of the galaxy sample, survey mask
and photo-z. In Sec.~\ref{sec:corrfunc} we describe our angular correlation
measurements including the error estimates and the impact of star
contamination. Sec.~\ref{sec:results} refers to our redshift space distortion
analysis and the implications for the growth rate of
structure. Sec.~\ref{sec:bao} is dedicated to discuss the evidence for the
 baryon acoustic feature in the measured correlation. Lastly,
 Sec.~\ref{sec:conclusions} contains our main conclusions.

\section{Data}
\label{sec:data}

\subsection{Selection of the Galaxy Sample}
\label{sec:sel}

We perform a color based selection of LRGs in the photometric catalog
of the final SDSS II data release
(DR7, \cite{2009ApJS..182..543A}). We follow
two main steps. The first one, based on that published 
by \cite{2006MNRAS.372L..23C}, aims at identifying the 
region in color--color space that is populated by high redshift 
LRGs (\cite{2001AJ....122.2267E}) by selecting all those objects that verify
\begin{eqnarray}
&&(r-i) > \frac{(g-r)}{4} + 0.36 , \nonumber \\
&&(g-r) > -0.72~(r-i) +1.7, 
\label{eq:selection1}
\end{eqnarray}
where the variables $g$, $r$, $i$ are {\it model} magnitudes corrected by extinction.
The second step is a set of cuts that are intended to minimize the star contamination in the sample,
\begin{eqnarray}
 && 17 < petror        < 21 , \nonumber\\
 && 0  < \sigma_{petror} < 0.5, \ \ \ \nonumber\\
 && 0  < r-i      < 2 ,\nonumber \\
 && 0  < g-r      < 3 , \nonumber\\
 && 22 < mag_{50} \left[{\rm mag/arcsec^2}\right]< 24.5 ,
\label{eq:selection2}
\end{eqnarray}
where $petror$ is petrosian magnitude (corrected by
  extinction), $\sigma_{petror}$ is the error on $petror$ and $mag_{50}$ is the surface
brightness in magnitude $petror$ at half-light radius $r_{50}$ (the
radius containing 50\% of petrosian flux), $mag_{50}=petror + 2.5 log (\pi r^{2}_{50})$.
These cuts yield a total of $\sim 1.27 \times 10^{6}$ objects with
redshifts in the range $\sim 0.4$ to $\sim 0.65$.

From the set of cuts in Eqs.~(\ref{eq:selection2}) the first ones corresponds to our magnitude limits
. The next ones ensure that
colors correspond to a galaxy and effectively eliminates very few objects. Lastly the cut that is most effective,
that in $mag_{50}$, with an upper bound on 24.5 to ensure well
measured galaxies and a lower bound on 22 to eliminate bright
point-like objects (i.e. stars). We have found that eliminating objects with $mag_{50}>24.5\,{\rm mag/arcsec^2}$ leads to
no difference in the clustering signal. However eliminating objects
with $mag_{50}<22\,{\rm mag/arcsec^2}$ reduces the amplitude of clustering at large scales by large factors 
even though they represents a small percentage of the total
sample. Hence, we
next discuss the motivation for this cut in more detail
(further evidence for this effect is given in
Sec. \ref{sec:starcontamination}). 

\begin{figure}
\includegraphics[trim= 3cm 13cm 1cm 2cm, clip = true, width=0.47\textwidth]{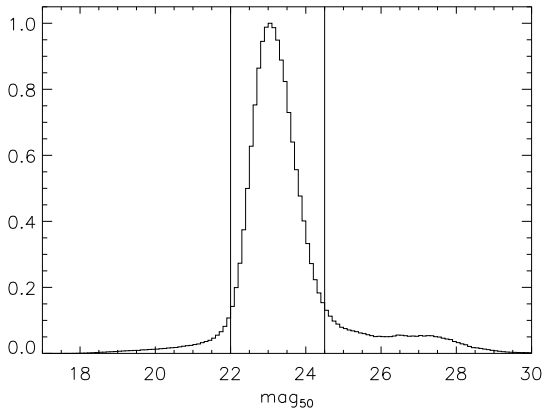}
  \caption{Histogram of surface brightness ($mag_{50}$) for all the
    objects in our catalog}
\label{fig:histomu50}
\end{figure}

\begin{figure}
\centering
\includegraphics[
width=0.46\textwidth]{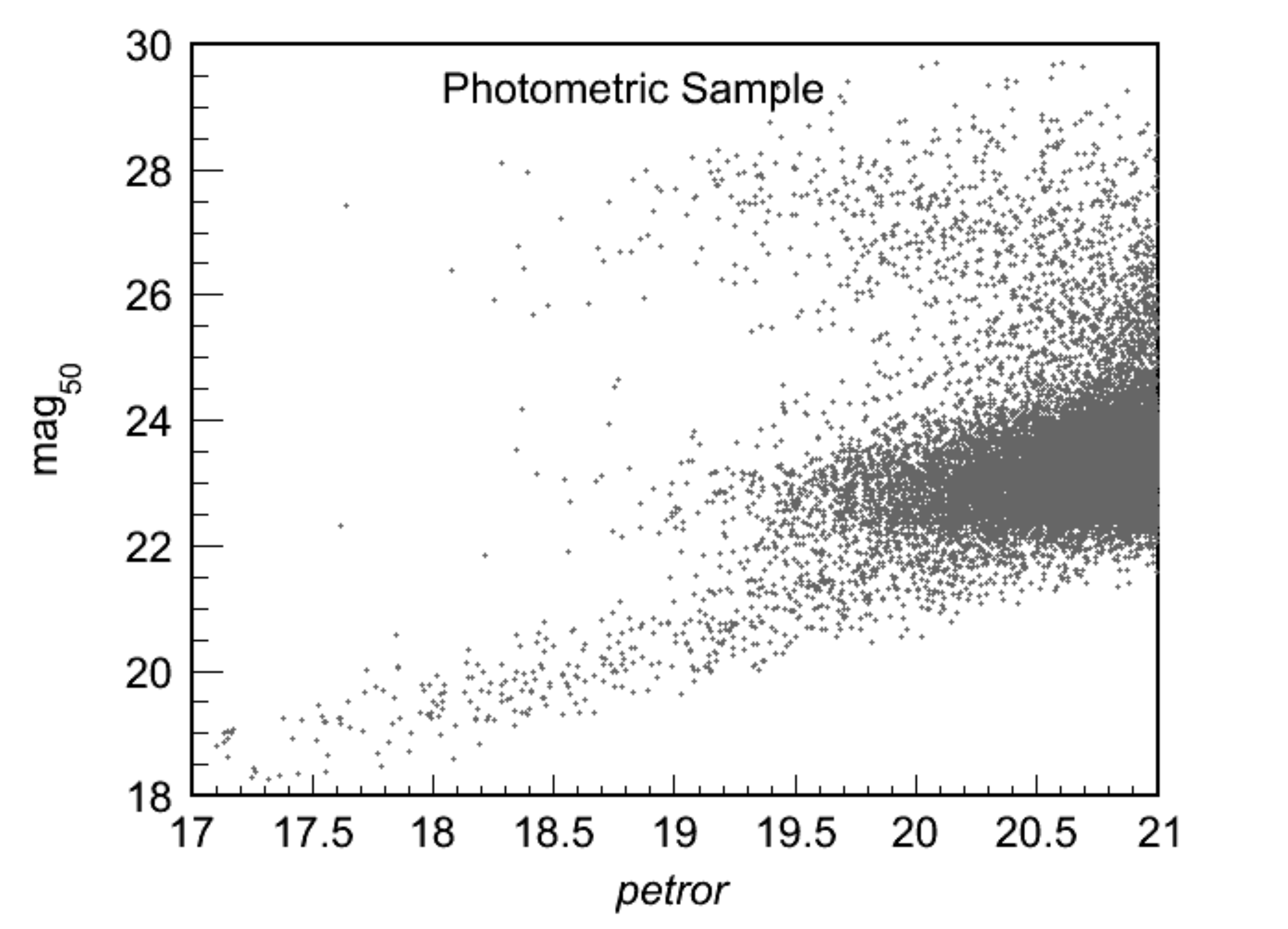}  \\
\includegraphics[
width=0.45\textwidth]{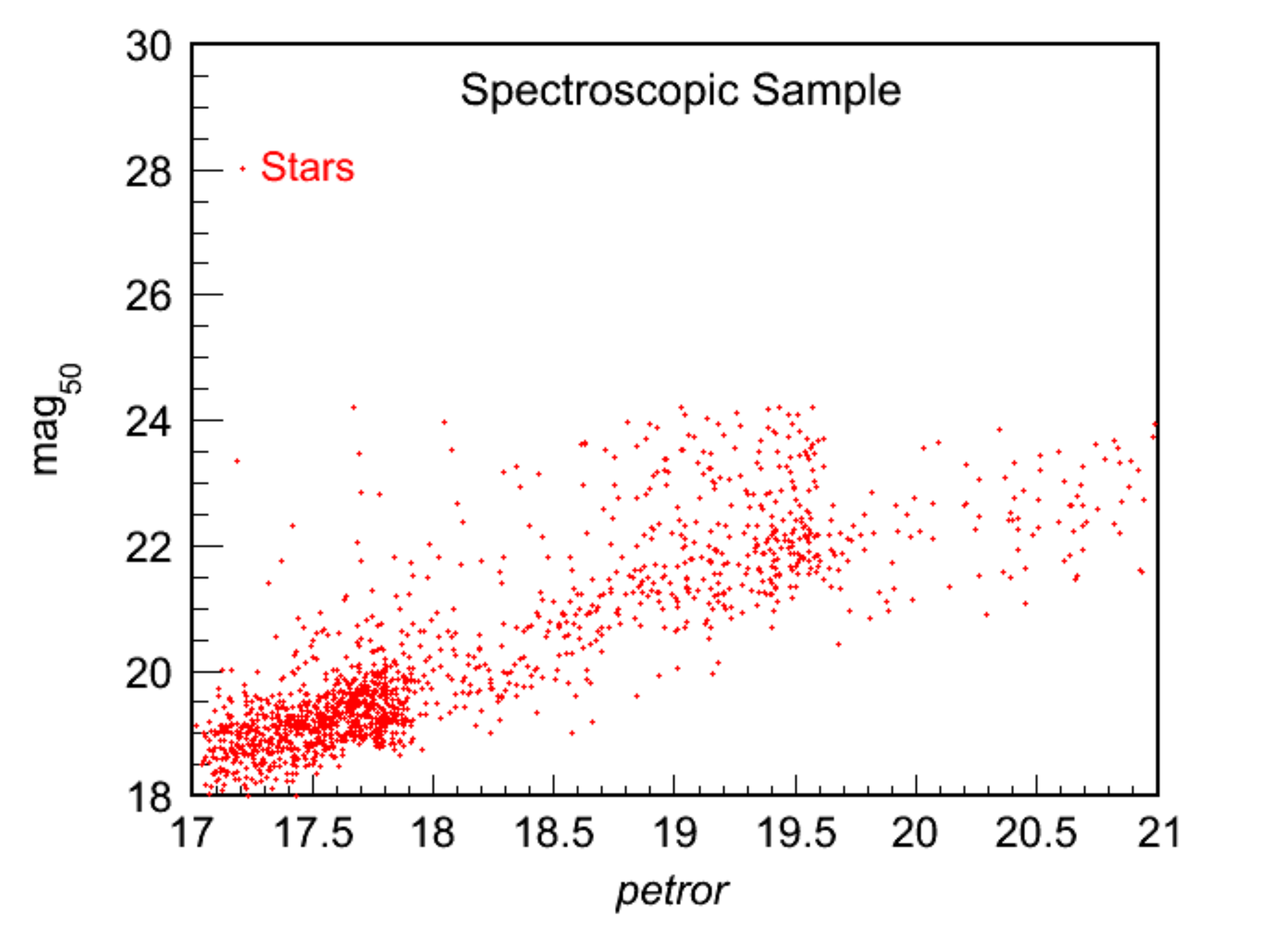} 
 \caption{Surface brightness ($mag_{50}$) vs. Petrosian $r$ apparent magnitude.
   Top panel corresponds to objects in our photometric LRG catalog, bottom panel
   to objects classified as stars in the SDSS DR7 spectroscopic sample that verifies the same
   selection as the top panel. Top panel shows some distinctive trends for
   $mag_{50}<22\,{\rm mag/arcsec^2}$ and for $mag_{50}> 24.5\,{\rm mag/arcsec^2}$, which contaminate our
   sample and can modify the clustering signal. Bottom panels makes
   clear that the region of $mag_{50}<22\,{\rm mag/arcsec^2}$ is populated by stars. The
   region with $mag_{50}> 24.5\,{\rm mag/arcsec^2}$ correspond to badly measured galaxies (not
   LRGs) and have no impact in our clustering analysis.}
\label{fig:histomu501}
\end{figure}

\begin{figure}
\includegraphics[
width=0.51\textwidth]{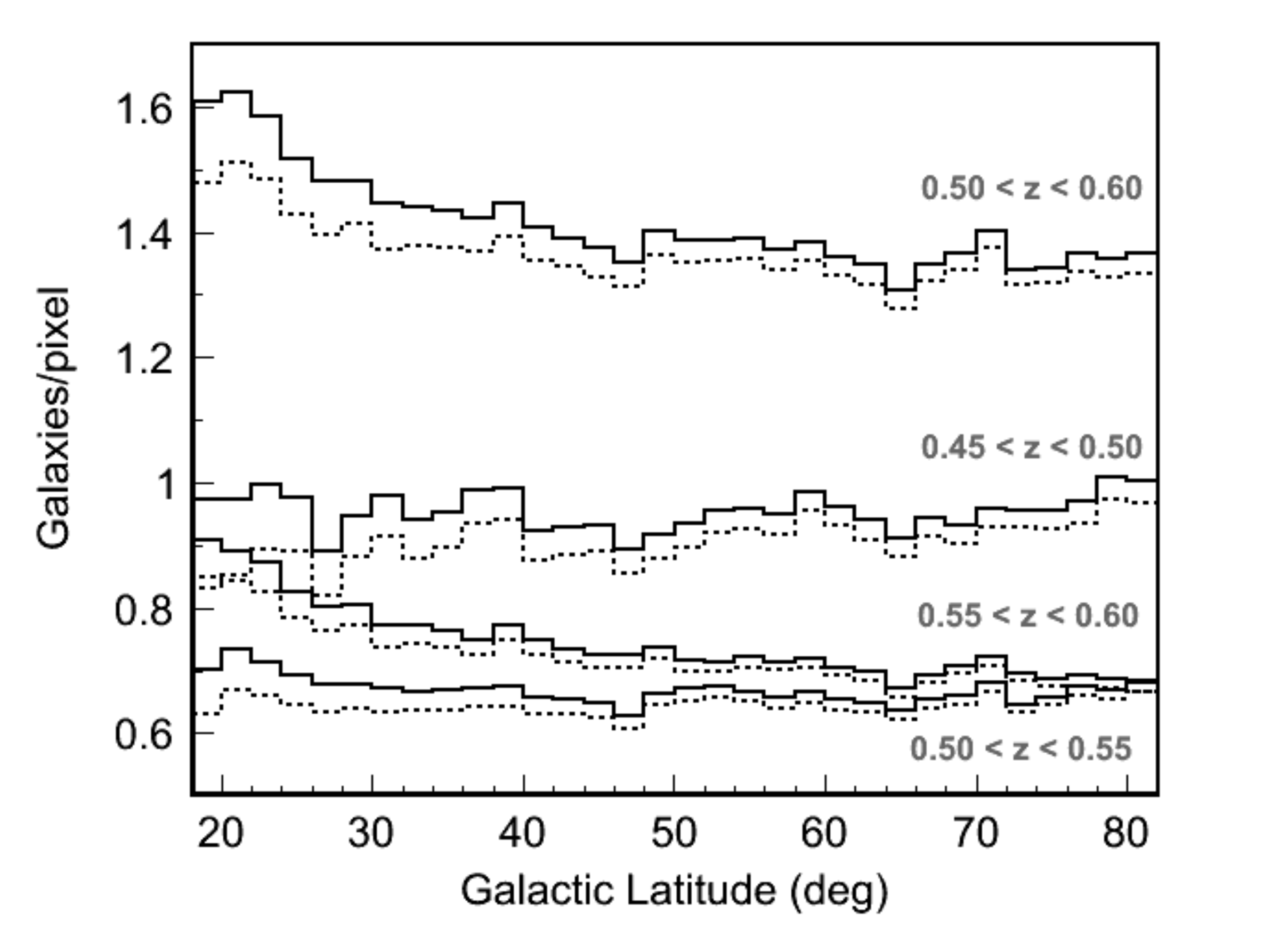}
  \caption{Histogram of number
    of galaxies per pixel (pixel size of 0.01 sq deg) as a function of
    galactic latitude for different slices in redshift as indicated in
    the figure. In solid line we plot the histograms when we select
    galaxies with $18<mag_{50}<24.5$, while in dashed-dot we cut
    $22<mag_{50}<24.5$. Galaxies with low $mag_{50}$ are contaminating
    low galactic latitudes ($b\lesssim 25^\circ$).}
\label{fig:galacticlatitude}
\end{figure}

The distribution of $mag_{50}$ is given in Fig.~\ref{fig:histomu50}. It is well
concentrated around $mag_{50}\sim 23\,{\rm mag/arcsec^2}$ but shows long tails due to
objects contaminating the LRG sample.
This contamination is more clearly depicted in the $petror$ vs. $mag_{50}$
diagram in Fig.~\ref{fig:histomu501}. Top panel corresponds to our
 photometric sample and shows a different trend for
 $mag_{50}<22\,{\rm mag/arcsec^2}$ and $mag_{50}>24.5\,{\rm mag/arcsec^2}$, with the core of LRGs lying in
 between. Bottom panel shows the same diagram but for the SDSS DR7
 spectroscopic sample, after imposing the selection in
 Eq.~(\ref{eq:selection1}).
This panel 
nicely shows that all objects with $mag_{50}<22\,{\rm mag/arcsec^2}$ could be contaminated
by stars. 
In addition, Fig.~\ref{fig:galacticlatitude} shows a histogram of number of objects
per pixel (here the pixel size is $0.01 {\rm deg}^2$) as a function of
galactic latitude and different redshift bins, including (solid) or excluding (dashed) 
galaxies with low $mag_{50}$. Objects with low $mag_{50}$ clearly concentrate at
low galactic latitudes introducing artificial density gradients
towards the galactic plane (which then results in large density
fluctuations at large scales). There is a slight gradient residual after imposing
the cut in $mag_{50}$, which we avoid but leaving out galactic latitudes (denoted
$b$) lower than $25^\circ$. We require, 
\begin{equation}
b \ge 25^\circ.
\label{eq:selection3}
\end{equation}
This yields a reduction of $\sim 3\%$ of the SDSS area used. 

\subsection{Angular Mask}
\label{sec:angularmask}

We built the angular mask using a {\tt Healpix} pixelization
(\cite{2005ApJ...622..759G})\footnote{http://healpix.jpl.nasa.gov} over the
entire sky with
$N_{side} = 512$ that yields a pixel size of $\sim 0.01$ ${\rm deg}^2$.

\begin{figure}
\centering{
\includegraphics[trim= 0cm 1cm 0cm 3cm, clip = true,
width=0.4\textwidth]{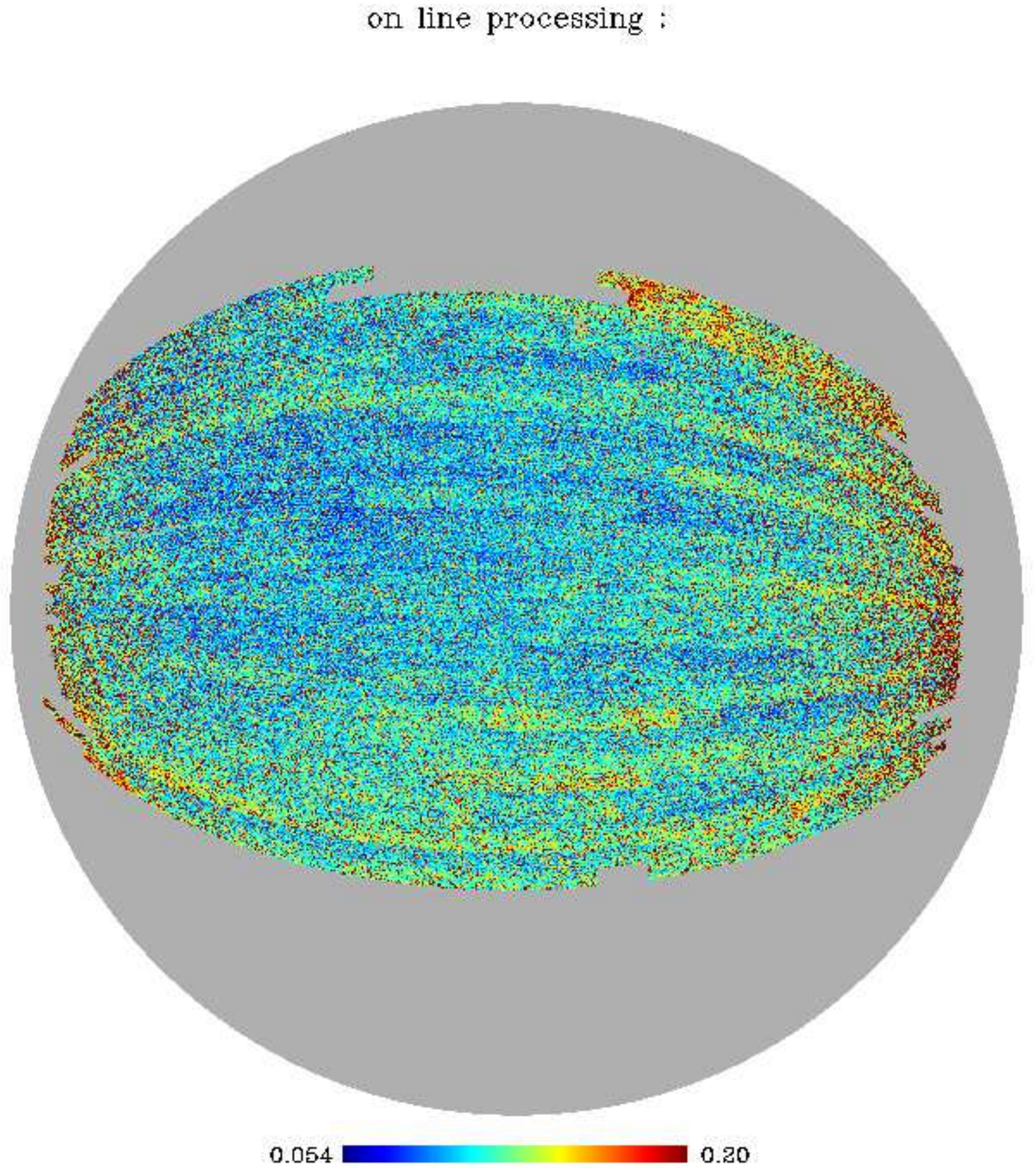}} \\
\centering{
\includegraphics[trim= 0cm 1cm 0cm 3cm, clip = true, width=0.4\textwidth]{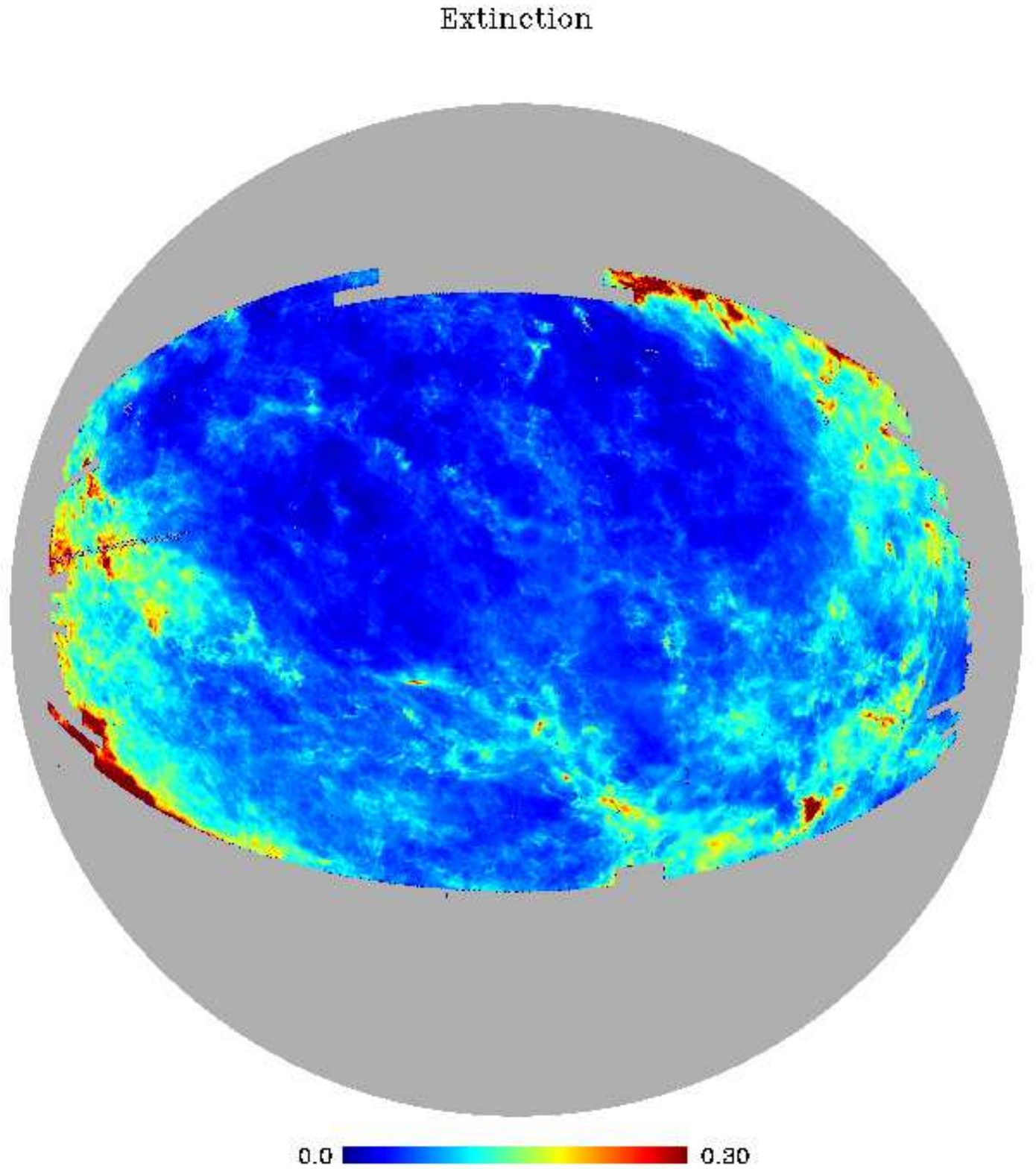}}
  \caption{Top Panel shows a map of mean error in $r$-band magnitude per
    (Healpix) pixel. Bottom Panel is the distribution of
    galactic extinction. Noticeably the mean error is larger in zones
    of higher extinction (see text for details).}
\label{fig:plotsextinction}
\end{figure}

We then eliminate from the analysis those pixels where
the geometric acceptance of the survey is compatible with bad or no
measurement by imposing a minimum number of galaxies per pixel
($N$/pixel) of 15. We notice that in order to build the mask we use {\it all} the objects in the
photometric catalog (i.e. not limiting by $petror<21$) because the
density of LRGs is very low to allow a robust and well pixelized mask construction.
In addition we look at galactic extinction and magnitude errors maps in order
to mask badly observed regions. Figure \ref{fig:plotsextinction} shows the distribution of errors
in r-band magnitude averaged in every pixel (i.e. mean error per
pixel) in the top panel, and in galactic
extinction in the bottom panel. There is a clear correlation between
these two quantities in regions of high extinction. 
Hence we suppress from our mask pixels with bad mean error rather than
applying the cut directly to the LRG selection, as this would imply introducing
artificially low density regions and corresponding systematic effects.
In summary we discard pixels with extinction higher than $0.2$ and mean
error higher than $0.3$. We have checked that using different pixelization sizes for the
mask, as well as different levels of acceptance of a pixel into the
mask (varying the threshold in extinction, mean error and $N$/pixel)
does not change appreciably the measured angular correlation.

\begin{figure*}
\centering
\includegraphics[trim= 1cm 6.7cm 1cm 3cm, clip = true,width=0.49\textwidth]{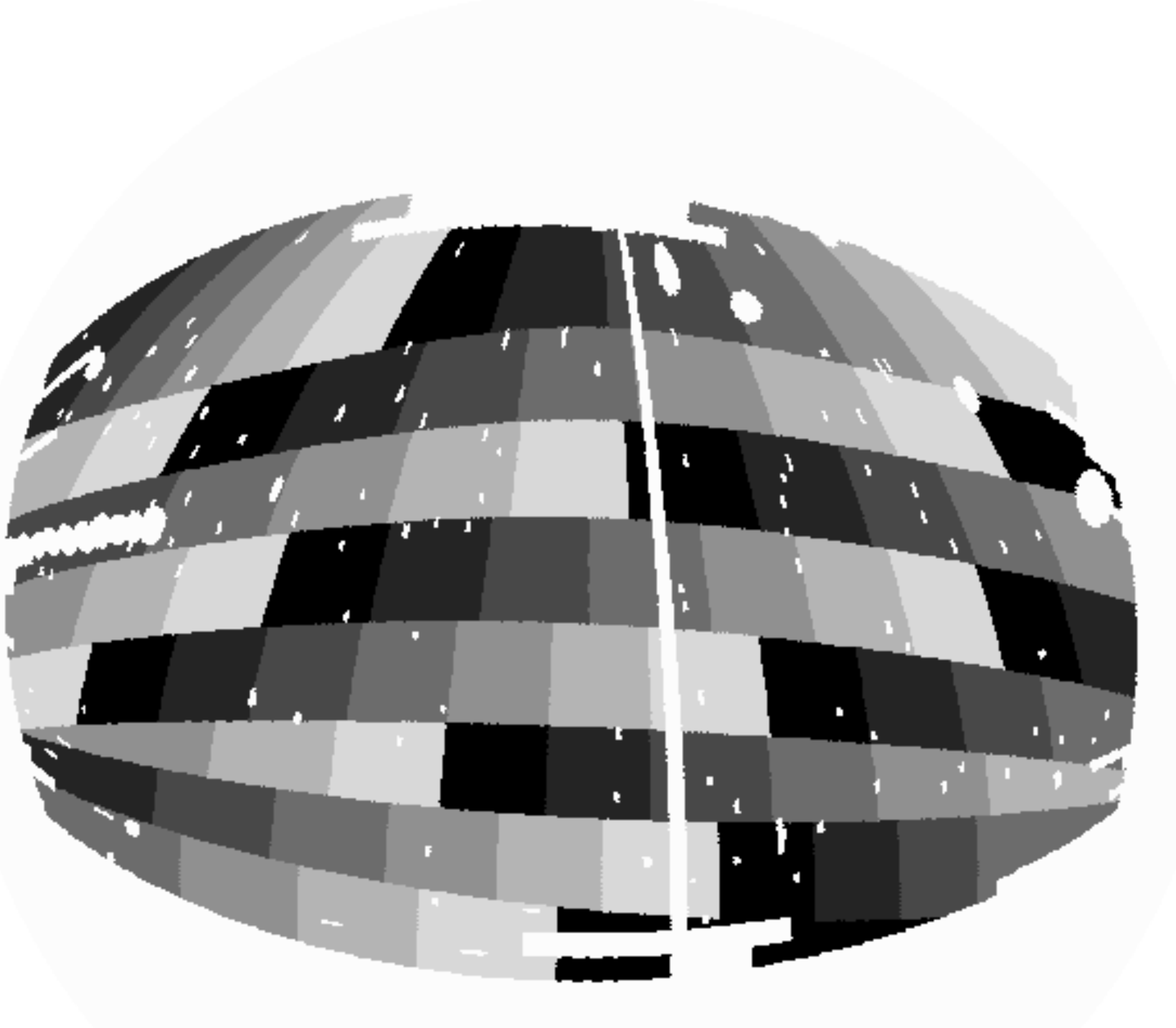}
\includegraphics[trim= 1cm 7cm 1cm 3cm, clip = true,width=0.49\textwidth]{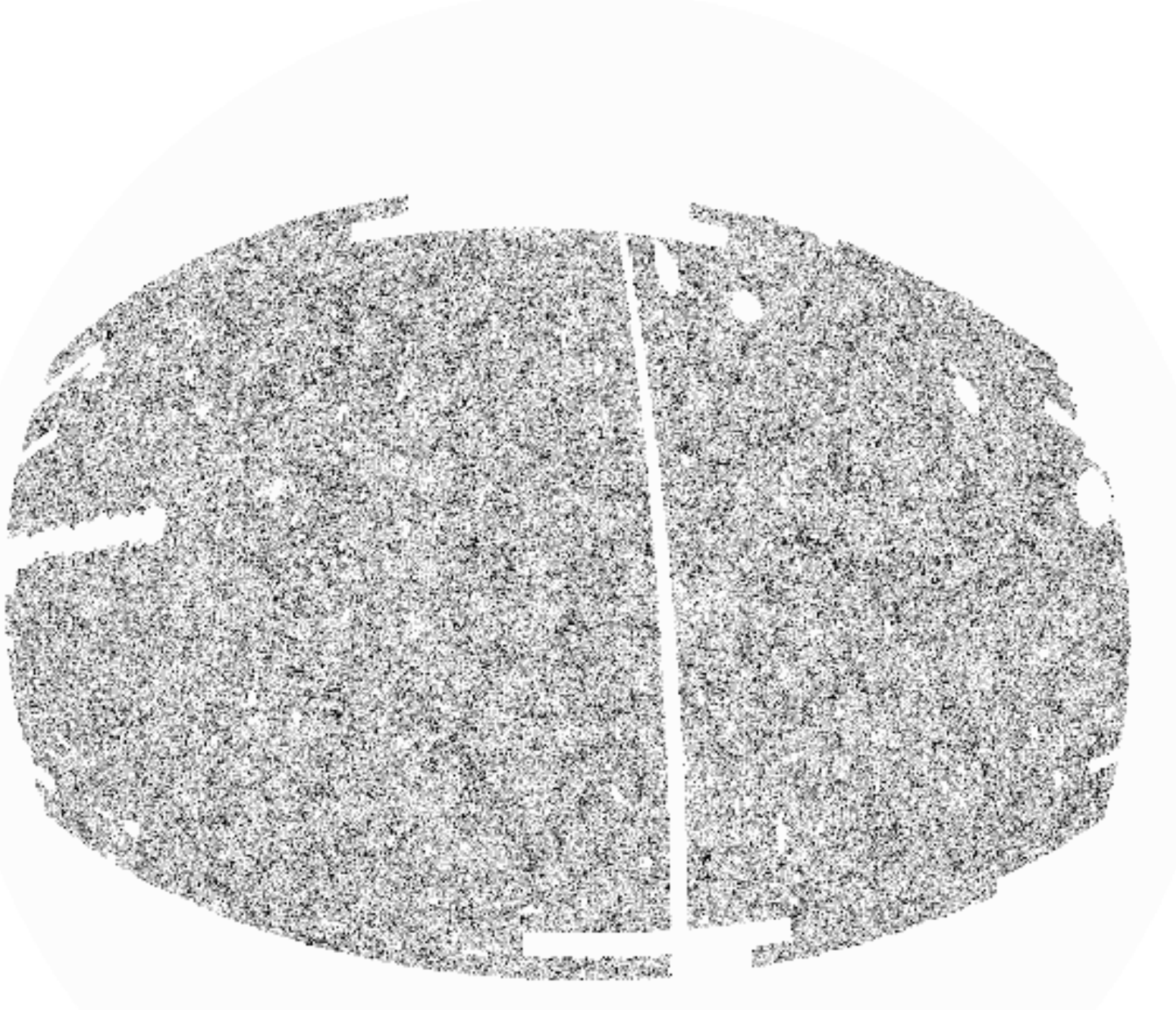} \\
\hspace{9cm}
\includegraphics[trim= 0cm 1cm 0cm 23cm, clip = true,width=0.45\textwidth]{LRGz0p5-z0p6-bgt20.pdf} 
\caption{Left Panel: The mask used in this analysis depicted with 
the {\tt Healpix} mollview projection routine. The white
  regions are excluded from the analysis.
  Different grey levels display the 81 JK zones
  used as one of two methods to estimate the errors. The vertical band is due to the photo-z used in
  this analysis. Right Panel: Angular density map of the galaxy distribution in the
  photometric bin $\left[0.5-0.6\right]$.}
\label{fig:mask}
\end{figure*}

The resulting angular mask is depicted in left panel of Fig.~\ref{fig:mask} in spherical
{\it equatorial} coordinates, with right
ascension ($ra$) and declination ($dec$) along the $x-y$ axis
respectively. It spans from $\sim 110^{\circ}$ to $260^{\circ}$ in
$ra$ and $75^{\circ}$ of $dec$ almost fully in the
norther hemisphere.
The vertical band at $ra \sim 172^{\circ}$ is due to the photo-z used in this
analysis (C. Cunha, private communication), that is described in the next section. 
Notice that we  only considered the largest contiguous area of the
survey, discarding stripes 76, 82 and 86 that contribute only a small
fraction of the total SDSS area.

This mask covers a total area of 7136 square degrees. We have verified that this angular mask is valid in the
full redshift range used in this paper, and is therefore used in all the analysis presented here.

Before moving onto the next section we include in the right panel of Fig.~\ref{fig:mask} the
resulting galaxy density map in the same coordinates as the mask and with the same healpix
pixelization ($N_{grid}=512$). As expected, it looks homogeneous over
the whole area. 

\subsection{Photo-z and Redshifts Distributions}
\label{sec:photoz}

In this paper we use the value added photometric catalog of
\cite{2009MNRAS.396.2379C} \footnote{Available at http://www.sdss.org/dr7/products/value$\textunderscore$added} that is based on the
Photoz2 tables of the full SDSS DR7 sample presented in \cite{2008ApJ...674..768O}.

This value added catalog was built using and extending the weighting
method technique of \cite{2008MNRAS.390..118L}.
As discussed in \cite{2008MNRAS.390..118L} and
\cite{2009MNRAS.396.2379C} the technique aims at estimating the
redshift distribution for a photometric galaxy sample (or selected subsamples)
rather than estimating individual galaxy redshifts. Hence, as an added
value the catalog provides accurate estimates of the redshift probability distribution,
$p(z)$, of each galaxy.

We applied photo-z quality cuts to the catalog in order to
remove badly defined $p(z)$ (e.g. double or multiple peaked
distributions that can represent
outliers) as well as very broad ones (that can be interpreted as galaxies with bad photo-z).
To this end we impose two cuts, $|z_{peak} - z_{mean}|<0.05$ and
$\sigma_z < 0.1$, where $z_{peak}$ is the peak of the distribution, $z_{mean}$ is computed as $\int
z \, p(z) \, dz$ and $\sigma_z=\int (z -z_{mean})^2 \, p(z) \, dz$
\footnote{The catalog provides $p(z)$ in 100 bins between $z = 0.03$ and $1.47$,
  hence these integrals are sums over these 100 bins.}.
The first cut eliminates roughly  $\sim 11\%$ of objects and the second
$\sim 9\%$. Imposed together these cuts reduce the sample by $\sim
16\%$. These threshold values in $\sigma_z$ and $|z_{peak}-z_{mean}|$
were obtained by identifying the tails in the distribution of values for these quantites
in the full catalog. One can of course be more conservative and impose
more stringent cuts but at the expense of biasing the population
towards brighter objects (that tipically have better photo-z) or
introduce shot-noise error due to large decrements in the number of
galaxies per bin. Our results, in terms of the $\chi^2$ of
the best-fit models that we obtain, are robust and stable in front of
the photo-z quality cut.

In this paper we split the galaxy sample into redshift bins 
according to whether the maximum of $p(z)$ lies in the bin or
not. That is, we identify the maximum of $p(z)$ as the photometric
estimate of the true redshift ($z_{phot}$) and do top-hat bins in photometric redshift.
In turn, one of the most important ingredients in order to interpret the galaxy
clustering signal is a robust estimate of the distribution in
true (spectroscopic) redshift, $N(z)$, of all the galaxies in each bin.
\cite{2009MNRAS.396.2379C} discuss several methods to obtain $N(z)$
and shows, using both mock SDSS catalogs and trainning spectroscopic
subsamples, that their best and almost unbiased estimate is provided
by the weighted sum of the training distribution (see also
\cite{2008MNRAS.390..118L}),
 which is equivalent to sum the $p(z)$ 
distributions of all the galaxies in the photometric bin,
\begin{equation}
N(z) = \sum_{i=1}^{N_{gal,bin}} p_i(z).
\label{eq:Nzestimator}
\end{equation}

\begin{figure}
\begin{center}
\includegraphics[width=0.4\textwidth]{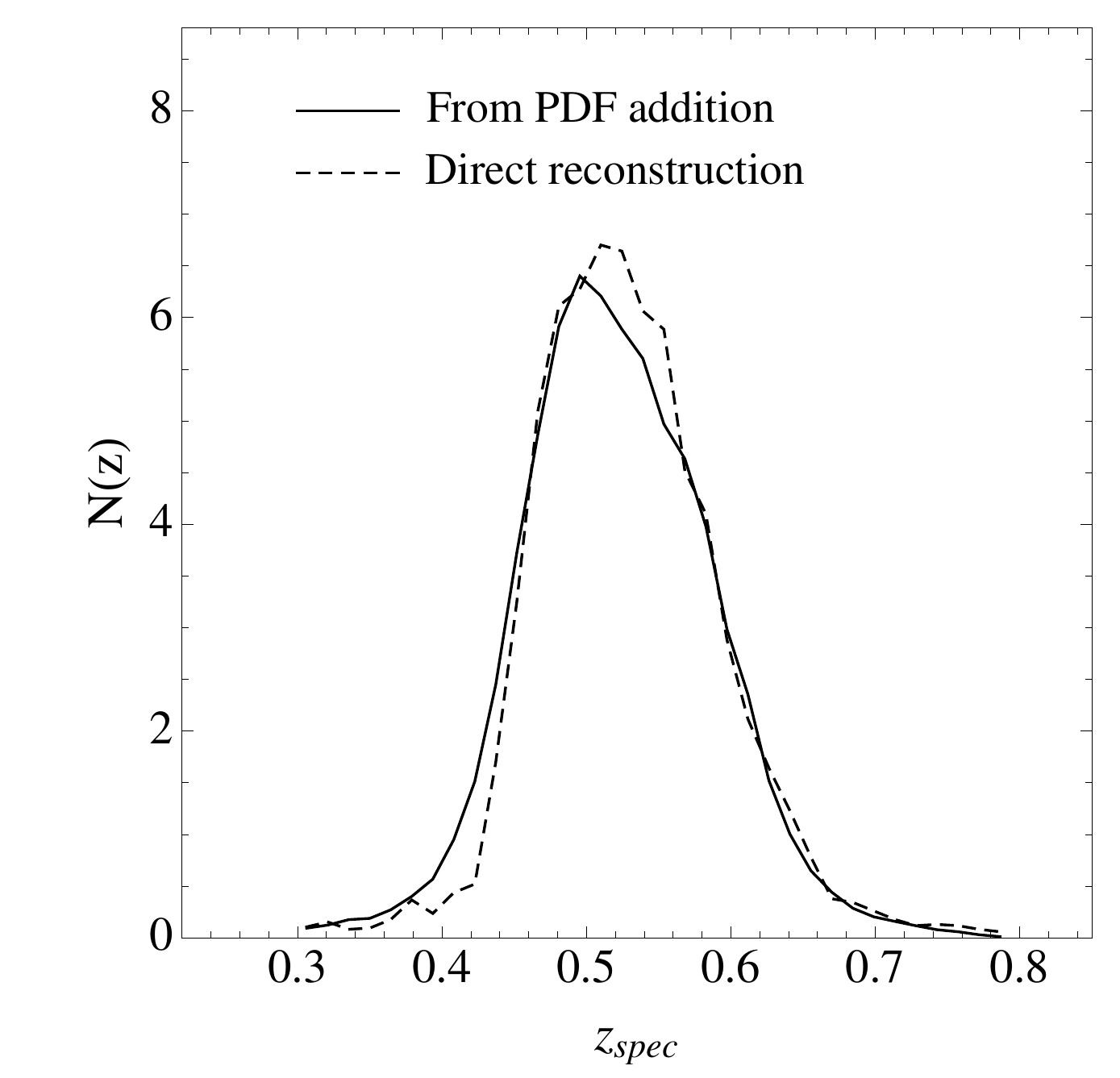} 
\caption{Direct reconstruction of the true redshift distribution for the
  spectroscopic sub-sample of our photometric catalog against the
  estimate for the same distribution using individual galaxy redshift
  probability distribution functions (PDF). The spectroscopic sub-sample was
  constructed from the 2SLAQ catalog, Cannon et al. (2006), while PDF's
  are provided as a part of the DR7 SDSS value added catalog of Cunha et al. (2009).}
\label{fig:dndzPDFvs2SLAQ}
\end{center}
\end{figure}

To test the accuracy in this determination of $N(z)$
we have selected all those galaxies in our sample that are also included in
the 2SLAQ spectroscopic catalog\footnote{2dF-SDSS LRG and
  Quasar survey, a stripe close to $0^{\circ}$ declination within the imaging area of DR7 (\cite{2006MNRAS.372..425C})} ($\sim 6000$
objects) and
computed their distribution of true redshifts as well as N(z) according to
Eq.~(\ref{eq:Nzestimator}). These two distributions are remarkably
similar as shown in
Fig.~\ref{fig:dndzPDFvs2SLAQ}. A
Gaussian fit to each of them shows that their peak differs by less than $1\%$
and their width by less than $9\%$ \footnote{Peak and
    width are defined as the mean and standard deviation of the
    best-fit Gaussian distribution.}. This difference is in perfect agreement with
the intrinsic scatter in true redshift distributions obtained from
different photo-z codes (e.g. see Table A1 in \cite{2011MNRAS.412.1669T}). Notice that
we can not use 2SLAQ to estimate $N(z)$ for our complete catalog since
our LRG selection is different from that in 2SLAQ (in particular the
magnitude cuts). Nonetheless the previous study shows the degree of
unknown in the red shift distribution. Hence we will use Eq.~(\ref{eq:Nzestimator}) to
estimate $N(z)$ for our red shift bins and will discuss in Sec.~\ref{sec:systematics} how our results vary when the width
and/or peak change by $9\%$ and $1\%$ respectively.

\begin{figure}
\begin{center}
\includegraphics[width=0.4\textwidth]{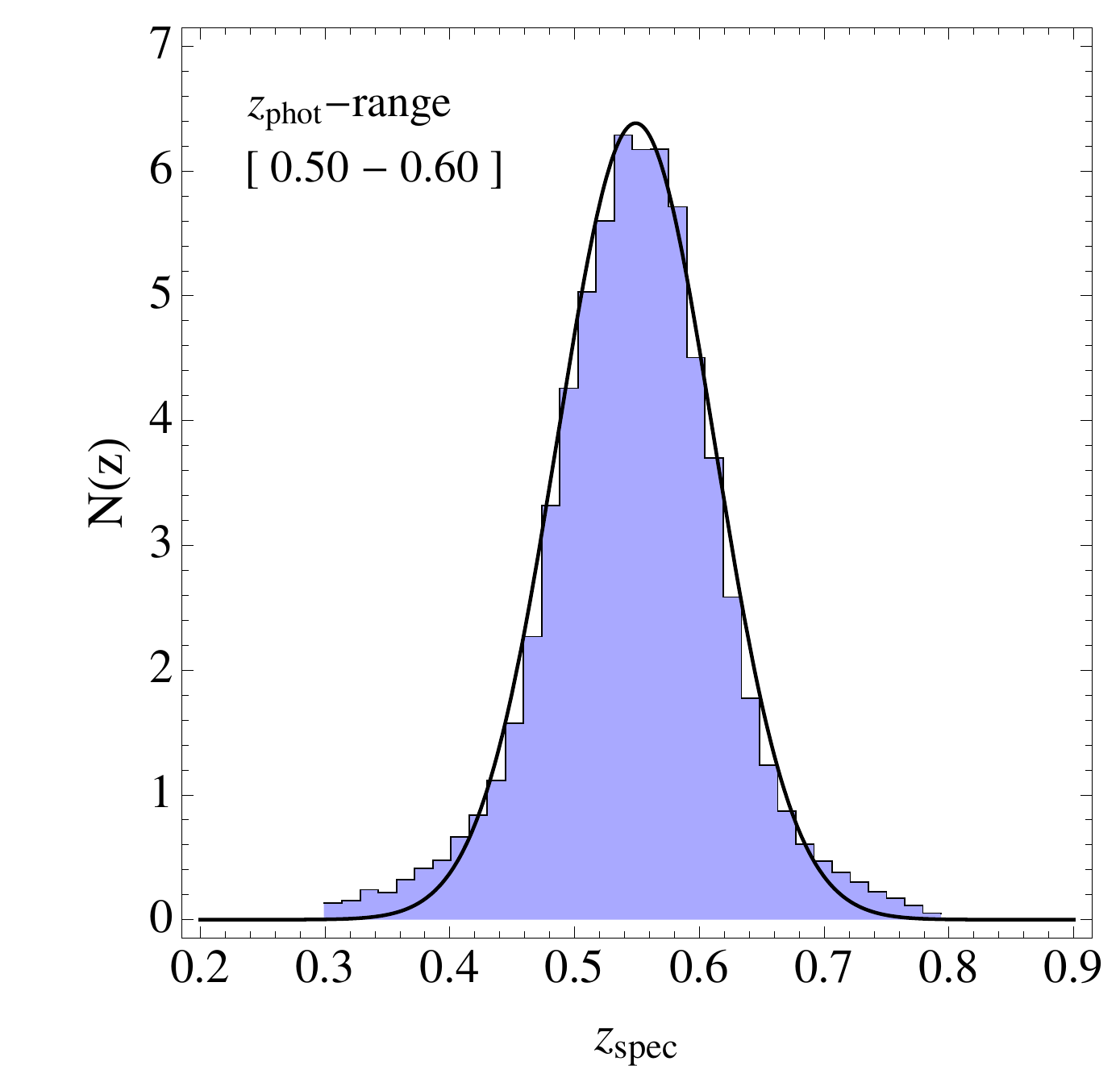} 
\caption{True (spectroscopic) redshift distribution for the bin $0.5-0.6$ resulting
  from sum of the individual redshift probability distributions. A fit to 
a Gaussian function (shown in solid black) yields a media of
$\mu=0.549$ and standard deviation $\sigma=0.062$.}
\label{fig:dndz0506}
\end{center}
\end{figure}

In this way, our selected sample of LRGs have a distribution in true red shifts that
peaks at $z \sim 0.5$ and extends roughly from $\sim 0.4$ to $\sim
0.65$. 
For our analysis we will mostly refer to a single top-hat photometric redshifts bin 
in the range $\left[0.5-0.6\right]$. 
Figure \ref{fig:dndz0506} shows the true
distribution of galaxies in this bin. The number of objects in this bin,
after the photo-z quality cut, is $664870$.
Notice that the bin width is
slightly larger than our typical photometric error at this redshift
($\sigma_z \sim 0.05$) hence choosing a narrower bin would yield
almost the same distribution of galaxies but at the expense of
increasing the noise in the measurement of $w(\theta)$ due to smaller number of particles in the bin
(see below).

\begin{figure}
\begin{center}
\includegraphics[width=0.4\textwidth]{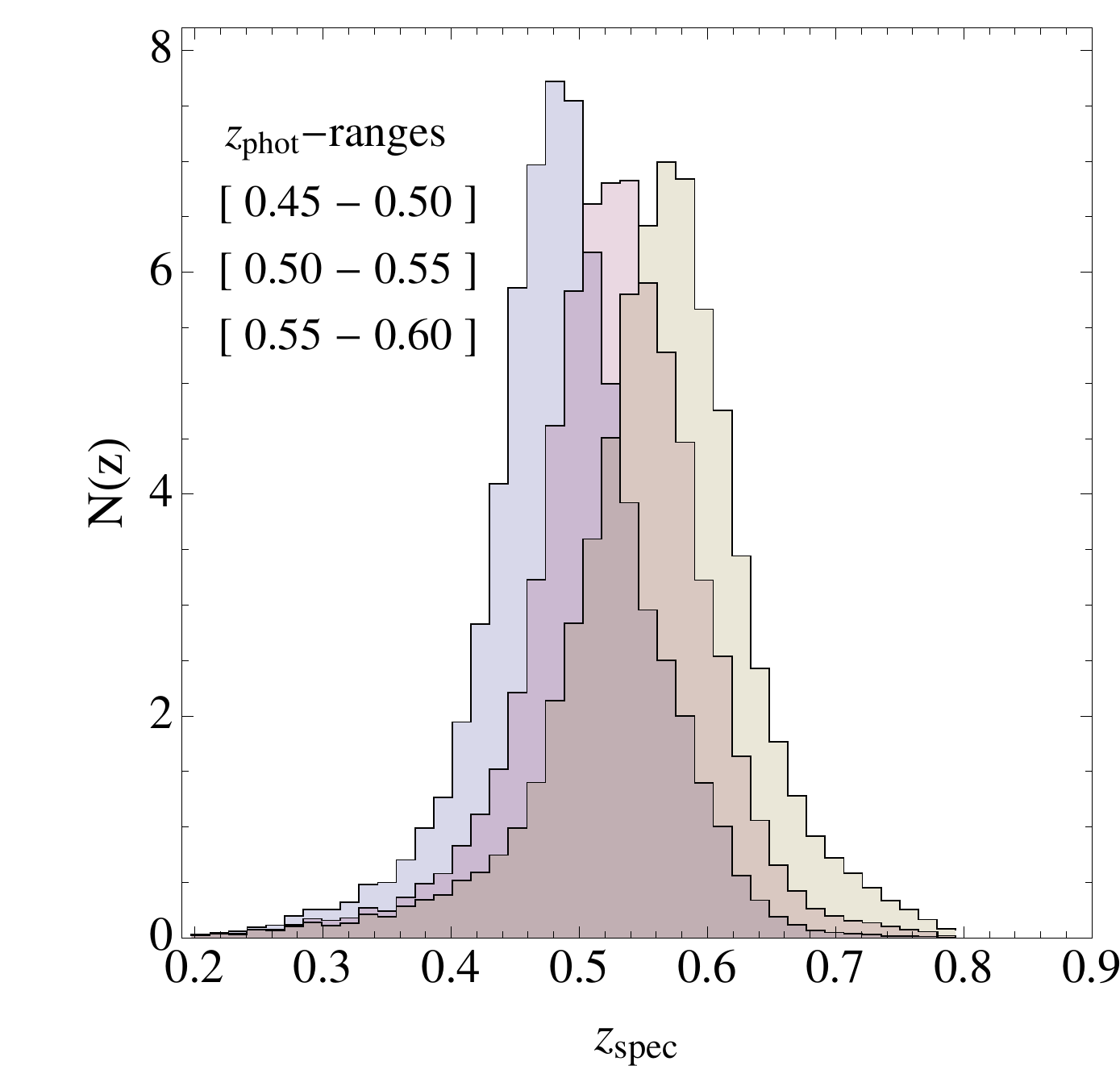} 
\caption{True redshift distribution for a set of ``narrow'' bins of
 width similar to the typical photometric error  ($\Delta z=0.05$).}
\label{fig:dndz_narrowbins}
\end{center}
\end{figure}

We also consider three narrower redshift bins,
with photo-z ranges $\left[0.45-0.5\right]$, $\left[0.5-0.55\right]$
and $\left[0.55-0.6\right]$. The redshift distribution for these cases
are shown in Fig.~\ref{fig:dndz_narrowbins}.
They are clearly highly correlated and not narrower than the wider
top-hat bin discussed above. The number of LRGs in these 3 bins
are $451753$, $317882$ and $346988$ respectively. In addition to an
increase of shot-noise and overlap, one expects the 
estimation of $N(z)$ to be not so robust for a bin narrower than the
intrinsic photo-z (a possible evidence for this is discussed in Sec.~\ref{sec:systematics}).
These are the reasons why we decided
to concentrate in a single redshift bin, and repeat our analysis in
these 3 bins as consistency checks. 

Redshifts bins lower than $z=0.45$ and higher than $z=0.6$ do not have enough number of LRGs to
obtain precise measurements of the angular correlation (we find
$52845$ LRGs in the photometric range $\left[0.4-0.45\right]$ and
$30412$ in $\left[0.6-0.65\right]$). At those extreme bins our
measurements also become too sensitive to ours cuts (e.g. in galactic
latitude and/or photo-z quality). This might be due to various reasons,
for example, large magnitude errors that correlate with galactic
latitude and lead to large photo-z errors. 

But perhaps the most worrisome issue happens at $z>0.6$ where we find 
a large extra-power over a broad range of large angular scales
(already for $\theta > 1^\circ$). Excess of power\footnote{We note
  that the exact meaning of excess is only loosely defined in the literature, in general is taken a
  roughly more than two sigma difference between model and measurements} on large scales has
already been found and discussed in different LRG selections based on 
SDSS DR5 in \cite{2009arXiv0912.0511S} and SDSS DR7 in
\cite{2011MNRAS.412.1669T,2010arXiv1012.2272T} (see also
\cite{2007MNRAS.374.1527B} and \cite{2007MNRAS.378..852P}). Nonetheless, we
only encounter this problem in the redshift bin $\left[0.6-0.65\right]$. 
We defer a discussion of possible reasons for Appendix~\ref{sec:appendix} and proceed
to discard these bins from our study hereafter.

\section{Correlation Functions}
\label{sec:corrfunc}

The angular correlation function measurements were performed 
starting from {\tt Healpix} angular maps as described in Sec.~\ref{sec:angularmask}  
($N_{side} = 512$, pixel size of $\sim 0.01$ ${\rm deg}^2$) and using
  a standard pixel estimator (\cite{2002MNRAS.333..443B}, \cite{2004ApJS..151....1E})
\begin{equation}
\hat{\omega}(\theta)=\frac{1}{N_{pairs}(\theta)}\sum_i\sum_j \delta_G^i \delta_G^j
\label{eq:pixelestimator}
\end{equation}
where $\delta_G^i=N_{gal}^i/\hat{N}_{gal} -1$ is the fluctuation in
number of galaxies in the i-$th$ pixel with respect to the mean in the
angular map, pixels $i$ and $j$ are separated by
an angle $\theta$ and $N_{pairs}(\theta)$ is the corresponding number of
pixel pairs. Pixels were weighted by 0 or 1 according to the angular
mask discussed in Sec.~\ref{sec:angularmask}. We have also implemented
a standard Landy \& Szalay \citep{1993ApJ...412...64L} estimator, and
the resulting measured correlations were within $1\%$ of that from
Eq.~(\ref{eq:pixelestimator}).

\begin{figure}
\centering
\leavevmode
\includegraphics[width=0.45\textwidth]{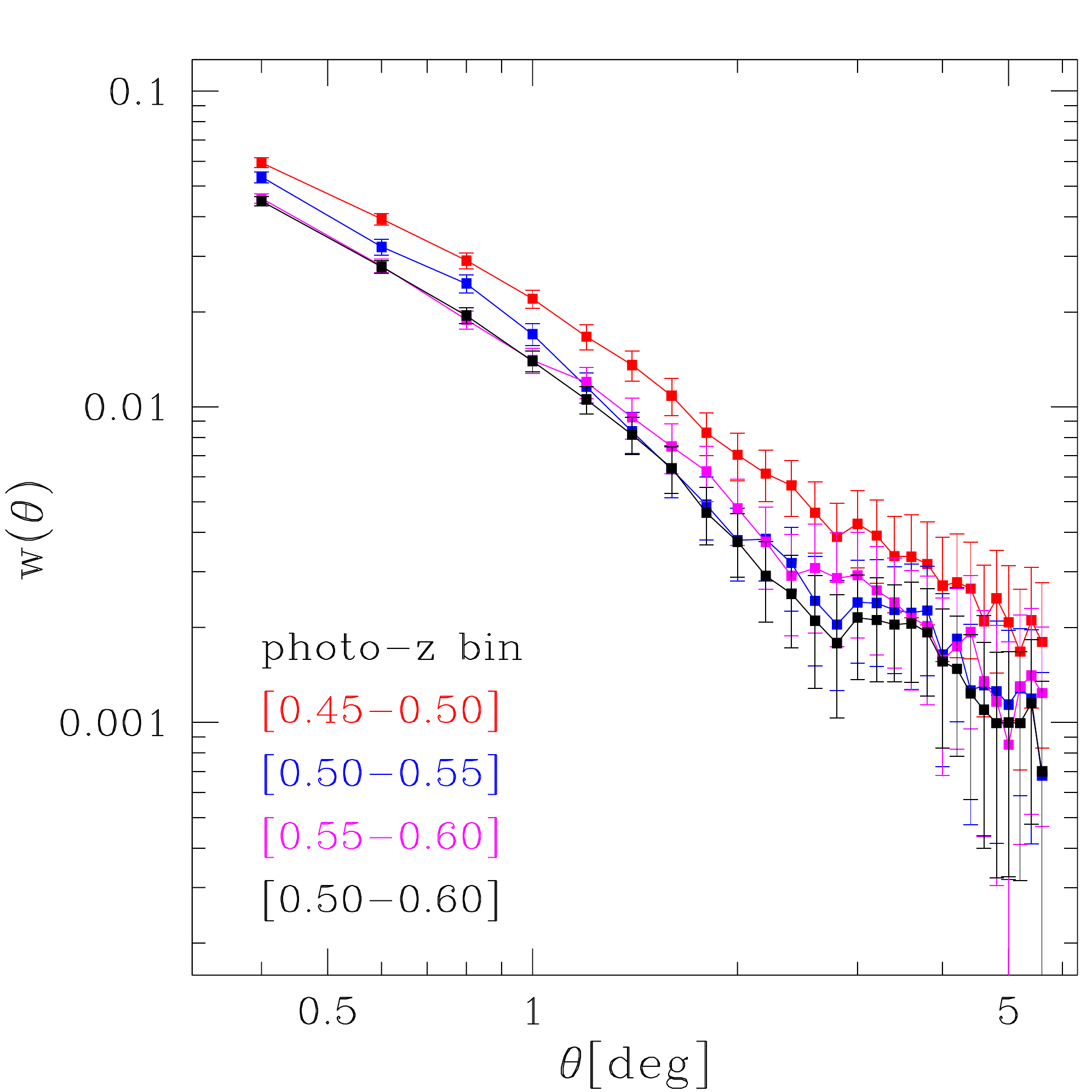} 
\caption{The measured correlation function in the different
 top-hat bins in photometric redshift. In this paper we focus on the
 bin [0.5-0.6] but test the robustness of our results throughout all
 the redshift range.}
\label{fig:plg}
\end{figure}

The measured correlations in the three bins of width $0.05$ as well as in the
bin $0.1$ are shown in Fig.~\ref{fig:plg}. Error bars displayed in this figure were obtained
using jack-knife resampling. In what follows we discuss our different
error estimates.

\subsection{Error Estimates}
\label{sec:errors}

We estimate the error and covariance between angular bins in the measured
correlation function using two independent methods. 

One method was implementing the standard jack-knife resampling technique, that to
date has been widely used in clustering analysis with correlation
functions (for a summary see \cite{2007MNRAS.381.1347C} and \cite{2009MNRAS.396...19N} and references therein).
To this end we divided the
angular mask in 81 jack-knife (JK) zones of similar area ($\sim 90 \,
{\rm deg}^2$ each) and
shape. These zones are shown in grey levels in Fig.~\ref{fig:mask}.
One then takes different realizations to be
all the sampled area except from one JK zone at a time. The 
covariance is then computed from the dispersion among the measurements
of $w(\theta)$ in the $N_{JK}=81$ resulting realizations,
\begin{eqnarray}
{\rm Cov}_{JK}(\theta,\theta^\prime)=\frac{N_{JK}-1}{N_{JK}}\sum_{i,j=1}^{N_{JK}}
(w^{(i)}_{JK}(\theta) - \hat{w}(\theta)) \nonumber \\ \times (w^{(j)}_{JK}(\theta^\prime) - \hat{w}(\theta^\prime))
\label{eq:covJK}
\end{eqnarray}
where $\hat{w}(\theta)$ corresponds to the full area, $w^{(i)}_{JK}$
to the $i$-th realization and the factor
$N_{JK}-1$ corrects from the fact that realizations are not independent. The positive aspect of the
jack-knife resampling is that this estimate is build out of the
data itself and hence it can account for systematic effects difficult to
capture otherwise. 
On the negative side is the fact that different realizations might share a large fraction of area. In principle
this is accounted for by the rescaling of the covariance in Eq.~(\ref{eq:covJK}), but  this JK resampling may not reflect
the true underlying statistical variance, particularly if the JK zones are too small or irregular

It is then desirable not to rely only on the jack-knife estimator.
Therefore we have also calculated the error and covariance matrix
using the analytical approach discussed in \cite{2011MNRAS.tmp..385C}
in which,
\begin{equation}
{\rm Cov}_{Th}(\theta,\theta^{\prime})=\frac{2}{f_{sky}} \sum_{\ell \ge 0}
\frac{2\ell+1}{(4\pi)^2} P_{\ell}(x) P_{\ell}(x^\prime) (C_\ell +
1/{\bar n})^2
\label{eq:covT}
\end{equation}
where $x=\cos(\theta)$ and $C_\ell$ is the analytical angular power
spectrum for the redshift distribution of interest. This estimate have
been extensively compared against ensemble errors drawn from simulated
photometric surveys assuming different binnings of the
data, survey depth, underlying photo-z, shot-noise contribution and
more, see \cite{2011MNRAS.tmp..385C}. 
In turn, \cite{2011arXiv1102.0968R} recently showed that the distribution of best-fit values
for the bias and growth rate of structure recovered in survey mocks agreed very
well with the errors obtained when using
Eq.~(\ref{eq:covT}). Furthermore, these
studies used an angular mask comparable to the simple geometry treated
in this paper.

To compute the angular spectra in Eq.~(\ref{eq:covT}) we assume a $\Lambda$CDM
cosmology with WMAP7 parameters (\cite{2011ApJS..192...18K}) 
and use the redshift distributions in
Figs.~\ref{fig:dndz0506}, \ref{fig:dndz_narrowbins}. 
In turn, initial values for large-scale bias $b$ and growth rate $f$ are obtained from a $\chi^2$
minimization using jack-knife errors  (see Sec.\ref{sec:results}). Provided with the full
$C_{\ell}$ spectra we then compute the covariance matrix in
Eq.~(\ref{eq:covT}) with $f_{sky}=0.1682$. 

\begin{figure}
\centering
\leavevmode
\includegraphics[width=0.4\textwidth]{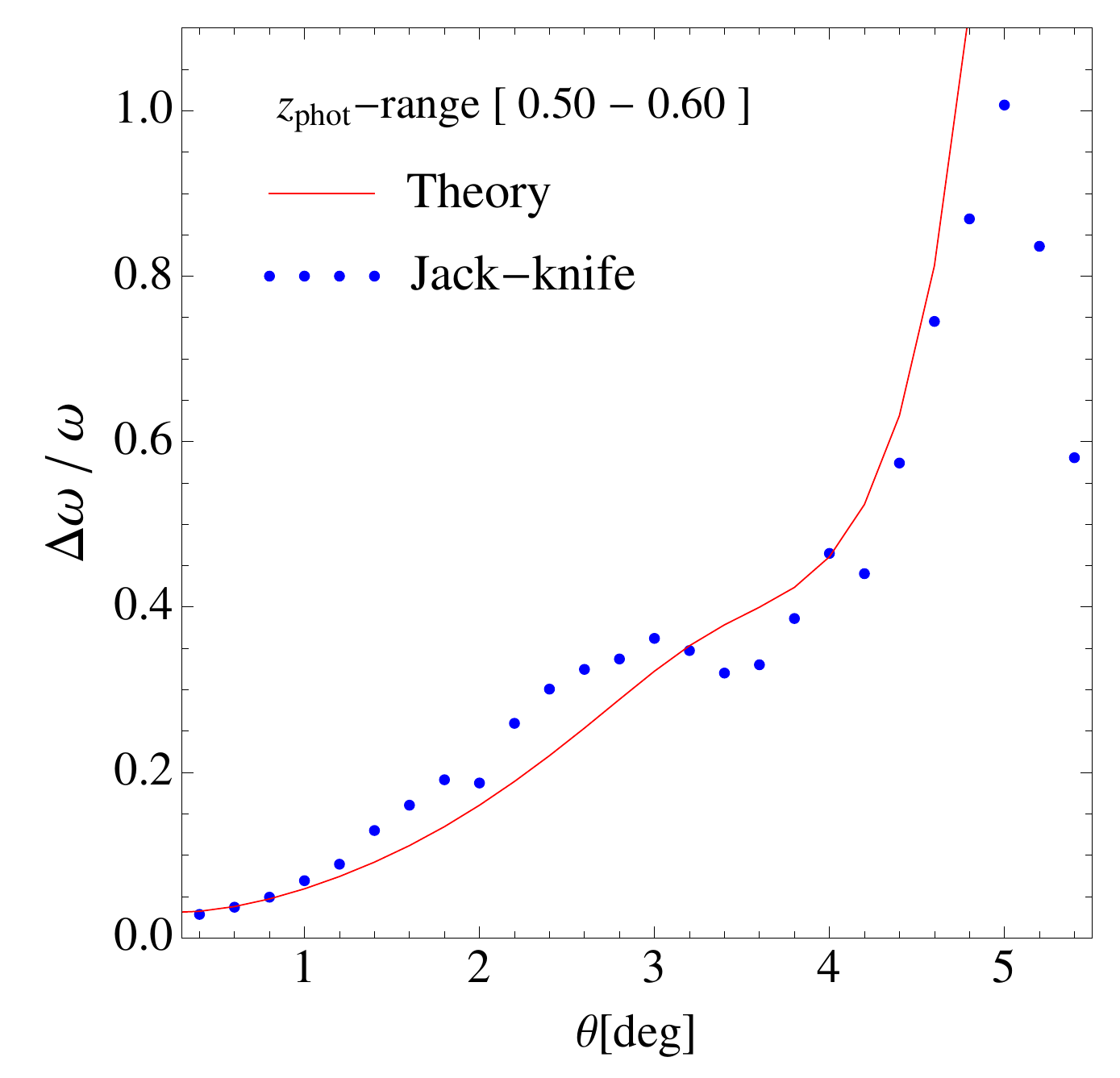} \\
\caption{Relative error in the measured correlation function. Symbols
  correspond to Jack-knife estimates using $81$ zones while
  solid line to the theoretical expressions in Crocce et
  al. (2011). In the later case the ratio is against the best-fit
  $w(\theta)$ model.}
\label{fig:werror}
\end{figure}

\begin{figure}
\centering
\leavevmode
\includegraphics[width=0.4\textwidth]{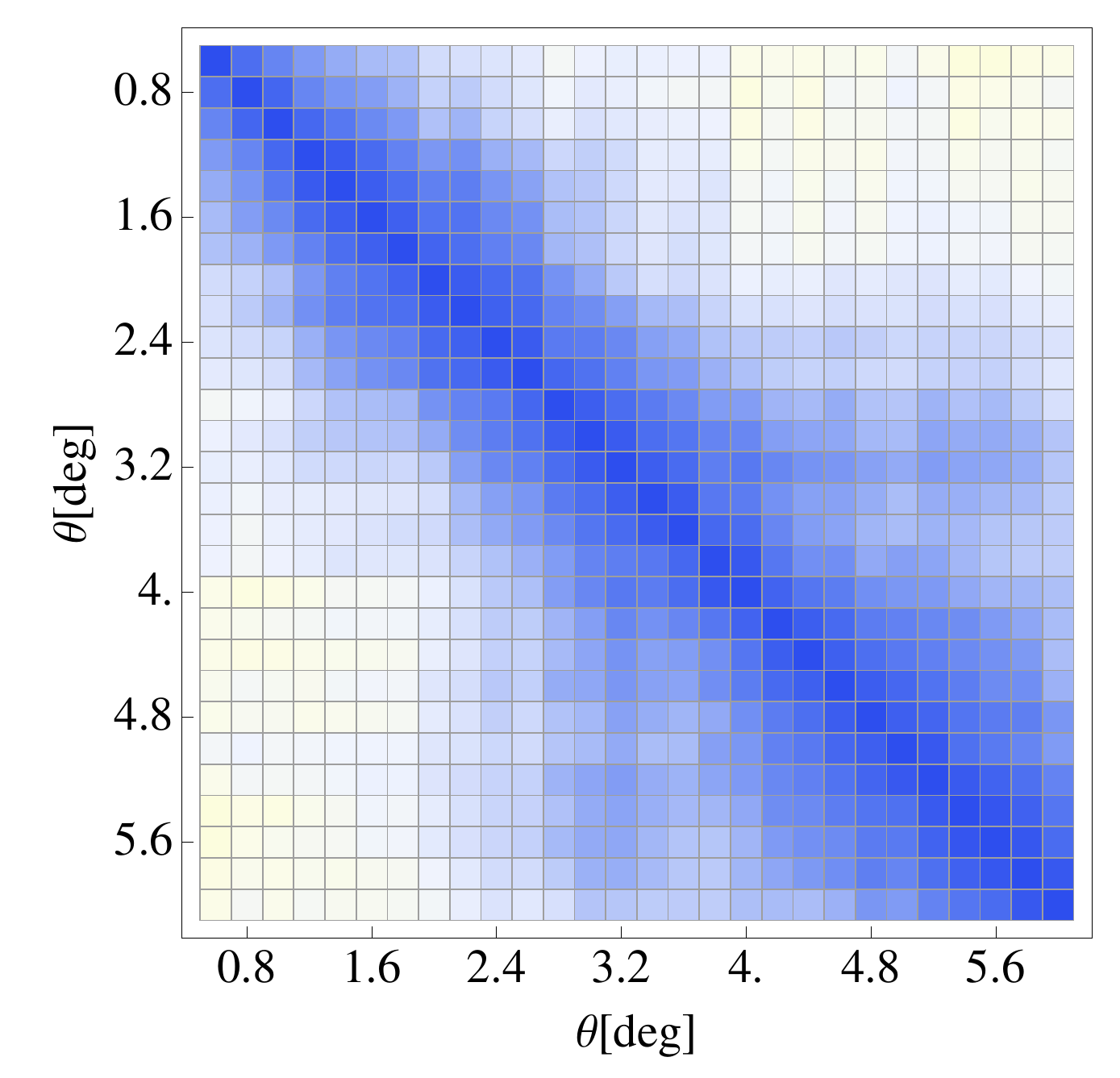} 
\includegraphics[width=0.07\textwidth]{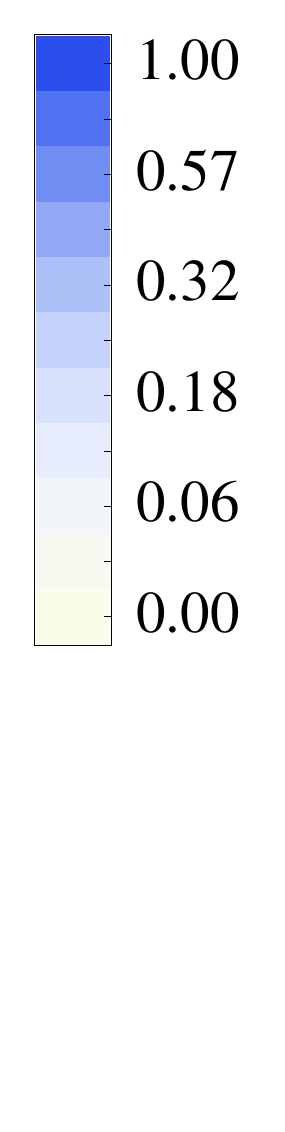} \\ 
\includegraphics[width=0.4\textwidth]{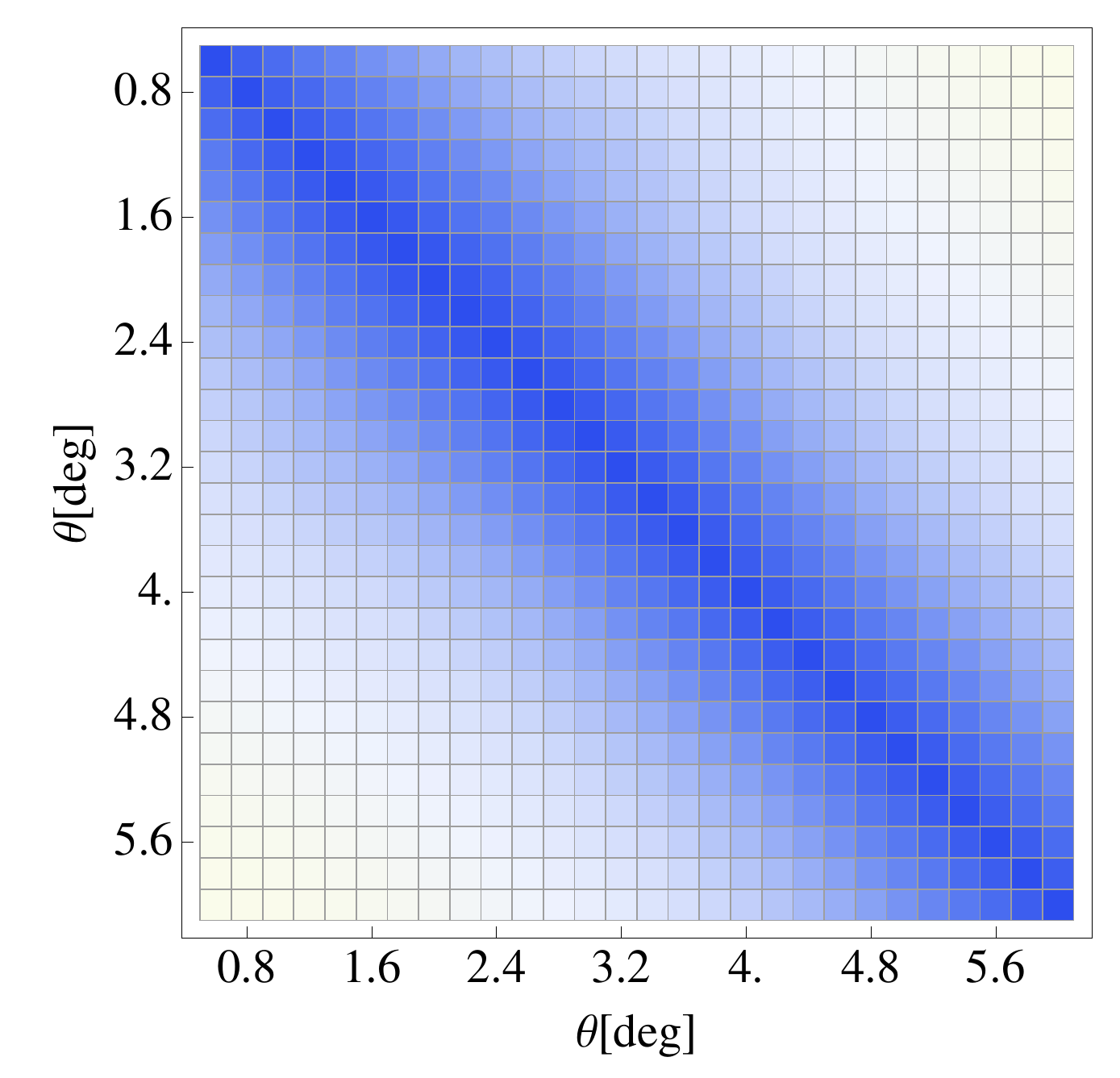} 
\includegraphics[width=0.07\textwidth]{cov-colorscale.pdf} \\ 
\caption{Reduced covariance matrix estimated with $81$ Jack-knife
  zones (top panel) or using the
  theoretical model in Crocce et al. (2011) (bottom panel).}
\label{fig:coverror}
\end{figure}

Figure \ref{fig:werror} shows the relative error $\Delta w / w$ in the
measurement of the angular correlation for our main case bin in
photo-z range $\left[0.5-0.6\right]$ (where $\Delta w \equiv {\rm Cov(\theta,\theta)^{1/2}}$). The agreement with the
relative error recovered from the jack-knife technique is remarkable 
. Figure \ref{fig:coverror} shows instead the reduced covariance matrix,
${\rm Cov}_{Red}(\theta,\theta^\prime)\equiv {\rm
  Cov}/\Delta w(\theta) \Delta w (\theta^\prime)$,
obtained from the data with jack-knife (top panel) and analytically (bottom panel)
They show a similar structure, with the jack-knife estimate being
more noisy as expected. In addition at large angles ($\theta \gtrsim 3^{\circ}$) there is a stronger 
covariance between separated angular bins in the jack-knife estimate probably due to systematics in the data.
Nonetheless, as discussed in Sec.~\ref{sec:results}, this has no major
impact in our study since the recovered best-fit models (and errors)
derived using either jack-knife or analytical estimates for the
covariance are in broad agreement. In all, the underlying reason why the jack-knife and analytical error estimates
coincide is due to the fact that the model correlation functions are in
good agreement with those measured in the data. This is the subject of
the forthcoming sections.

\subsection{Star contamination}
\label{sec:starcontamination}

Lastly, we study the star contamination of the sample as a function
of the redshift. The quality of the star-galaxy separation is a major
concern in photometric catalogs since the broadband colors of
stars can mimic
those of galaxies and yield similar photometric redshift estimates. 
and important distortions in the clustering signal.

We investigate the degree of contamination in our selected photometric sample
performing the same selection,
Eqs.~(\ref{eq:selection1},\ref{eq:selection2},\ref{eq:selection3}), 
in the SDSS spectroscopic sample. In addition we only
take the spectroscopic objects that overlap with our angular mask. 
For the bins where our
analysis is performed we find $f_{stars}=4\pm 1\,\%$. That is, a negligible dependence with
redshift and a broad agreement with
the residual contamination found in comparable clustering studies at
these redshifts (\cite{2009arXiv0912.0511S,2011MNRAS.412.1669T}).

\begin{figure}
\begin{center}
\includegraphics[width=0.4\textwidth]{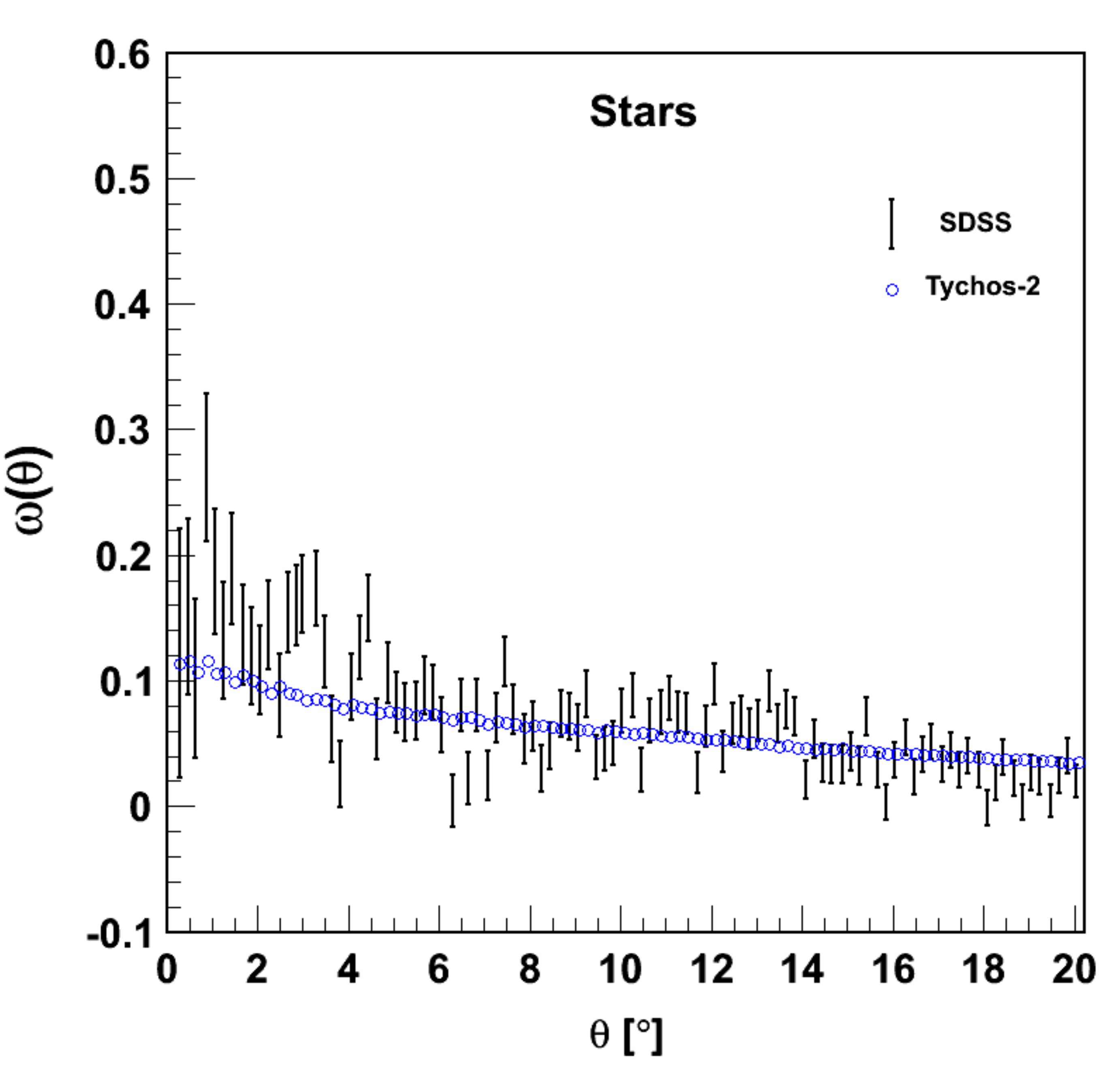} \\
\caption{Angular correlation function of stars from the SDSS spectroscopic
  sample (error bars) and the Tychos-2 catalog (circles) verifying our sample selection
  and mask\label{fig:stars}. The correlation is well fit by 
  $w_{stars,fit}(\theta)= 0.0904 - 0.00313 \,\theta$. Displayed error
  bars correspond to Poisson estimates, $\Delta w = (1 + w) /
  \sqrt{N_{pairs}}$ (negligible for Tychos-2).}
\label{fig:wtheta_stars}
\end{center}
\end{figure}

The next step is to estimate what is the impact of this contaminants in
the large scale angular clustering signal since they introduce a
density gradient through the galactic plane. Hence, we measure the correlation function of stars from the
SDSS spectroscopic sample, relaxing the cut in $mag50$ to have enough
statistics. We also use the publicly available Tychos-2 star 
catalog \citep{2000A&A...363..385H,2000A&A...355L..27H} cut to the
same selection and mask as our LRG sample to obtain a second estimate
of the correlation of stars. Both determinations
are in perfect agreement, as presented in Figure~\ref{fig:stars}, where the lines
represent the correlation function for the SDSS sample, and the circles correspond to the
Tychos-2 catalog.
This correlation
is then included in the theoretical model for $w(\theta)$ taking into account that LRGs and stars are 
uncorrelated populations, as (see also \cite{2006ApJ...638..622M},\cite{2007ApJ...658...85M}):
\begin{eqnarray}
w_{obs,model}(\theta,z) = (1-f_{stars})^2 w_{gal,model}(\theta,z) + \nonumber \\ f_{stars}^2
w_{stars,fit}(\theta) 
\label{eq:wstars}
\end{eqnarray}
where $w_{obs,model}$ is the model for the ``observed'' correlation function,
$w_{gal,model}$ is the model for the true correlation
function of galaxies, $w_{stars,fit}$ is a simple
fit to the measured correlation function for 
stars (see Figure~\ref{fig:stars}). 
Notice that in Eq.~(\ref{eq:wstars}) we have removed the explicit dependence of the star
fraction and correlation with redshift for simplicity.

\section{Bias and Growth of structure}
\label{sec:results}

In this section we will employ the measured angular correlation
function to place joint constraints in the growth rate of structure 
and bias of the LRG sample.

\subsection{Theoretical model}

We will use the theoretical model for the angular correlation function
presented in \cite{2011MNRAS.tmp..385C}. It was extensively tested against mock
 catalogs of photometric surveys with specifications similar
to our present case. It was shown that the inclusion of redshift-space
distortions and bias to linear order together with a model for
non-linear matter clustering accurately
reproduced the measured angular correlation in scales $\theta \gtrsim 0.5^\circ$
for redshift bins centered at $z\sim 0.5$. We now recall some
basic expressions of the model and refer the reader to \cite{2011MNRAS.tmp..385C} for 
further details. 

The model angular correlation function is given by,
\begin{equation}
w_{gal,model}(\theta) = \int dz_1 n(z_1) \int dz_2 n(z_2) \ \ \xi^s({\bf r}_{12})
\label{eq:wtheta}
\end{equation}
where ${\bf r}_{12}={\bf r}_{12}(z_1,z_2,\theta)$ is the separation of
a pair of galaxies at redshift $z_1$ and $z_2$ subtending an angle
$\theta$ with the observer. In turn, $n(z)$
is the spectroscopic redshift distribution of the photometric sample
under study. Notice that the inclusion of photo-z errors in the theory
is solely through $n(z)$ (e.g. \cite{2003ApJ...595...59B}).

\begin{figure}
\begin{center}
\includegraphics[width=0.4\textwidth]{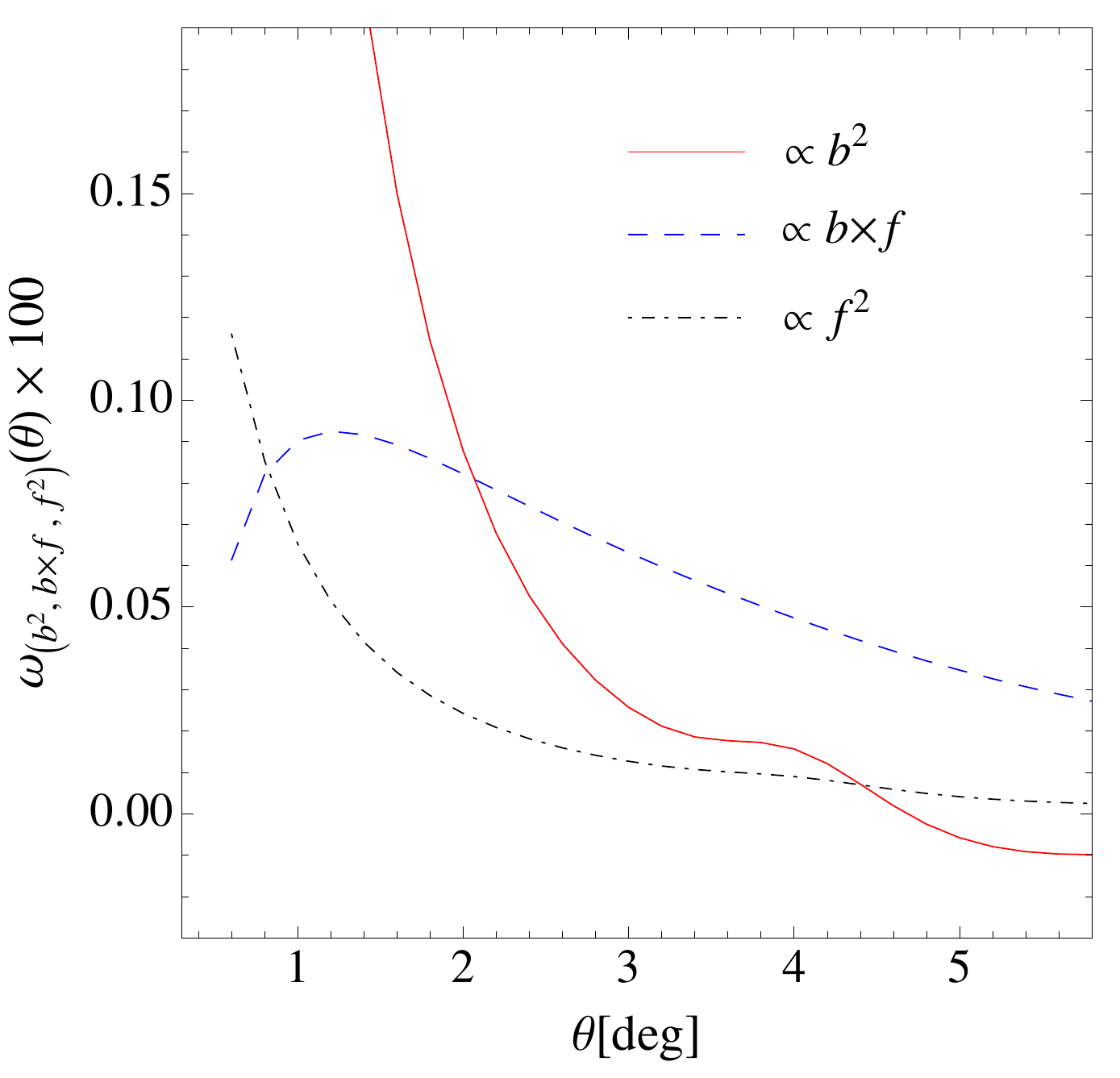} \\
\caption{Angular dependence of the three additive terms making up the model angular
  correlation. They are proportional to $b^2(z)$, $b(z)f(z)$ and $f^2(z)$ (in units of $\sigma_8(z)$)
and are shown in solid red, dashed blue and dot-dashed black respectively. Their different
  shapes make it possible to constrain simultaneously $b(z)$ and
  $f(z)$. This figure corresponds to our photo-z bin $\left[0.5-0.6\right]$.}
\label{fig:components}
\end{center}
\end{figure}

Given the area and mean redshift of our sample we are allowed to make
the plane-parallel approximation (see \cite{2010MNRAS.409.1525R}
  and references therein for discussions of its validity).
In this limit, the redshift space correlation is given by (\cite{1992ApJ...385L...5H})
\begin{equation}
\xi^s(s,\mu)=\xi_0(s) P_0(\mu) + \xi_2(s) P_2(\mu) + \xi_4(s) P_4(\mu)
\end{equation}
where $P_\ell$ denote the standard Legendre polynomials,
$s^2=r^2(z_1)+r^2(z_2)-2 r(z_1) r(z_2)\cos \theta$ and
 $\mu = (r(z_2)-r(z_1))/s$, with $r(z)$ being the comoving distance to redshift $z$.
The $\xi_\ell$ are the multi-poles of the spatial correlation
\begin{eqnarray}
&&\xi_0(r) = (b^2 + 2 b f/3 + f^2/5) \left[ \xi(r) \right] \nonumber \\
&&\xi_2(r) = (4 b f/3 + 4 f^2/7) \left[ \xi(r)-\xi^\prime(r) \right]    \nonumber \\
&&\xi_4(r) = (8 f^2 /35) \left[ \xi(r) + 5/2 \, \xi^\prime(r) - 7/2 \, \xi^{\prime\prime}(r)\right]
\label{eq:ximultipoles}
\end{eqnarray}
with $\xi^\prime \equiv 3\, r^{-3} \int_0^r \xi(x) x^2 dx$ and $\xi^{\prime\prime} \equiv 5\, r^{-5} \int_0^r \xi(x) x^4 dx$.
Hence the angular correlation in Eq.~(\ref{eq:wtheta}) can be written as,
\begin{eqnarray}
w(\theta) &=& (b^2 + 2 b f/3 + f^2/5) \, w_0(\theta) \, + \nonumber \\  
&& + \,(4 b f/3 + 4 f^2/7) \, w_2(\theta) + (8 f^2 /35) \,w_4(\theta) \label{eq:wtheta2} 
\end{eqnarray}
where $w_\ell(\theta)$ are the bin projection of the functions in square
brackets in Eq.~(\ref{eq:ximultipoles}).
Equation~(\ref{eq:wtheta2}) can of course be re-arranged into three terms
scaling as $b^2$, $b f$ and $f^2$,
\begin{eqnarray}
w(\theta) &=& b^2 w_0(\theta) +  b f \,(2/3 \, w_0(\theta) + 4/3 \, w_2(\theta)) + \nonumber \\ 
 && + f^2 (1/5 \, w_0(\theta) + 4/7 \, w_2(\theta) + 8/35 \, w_4(\theta)).
\end{eqnarray}
The fact that each of these 3 terms have a different angular dependence
(through the different linear combinations of $w_\ell$) makes it possible to
constrain both $b$ and $f$ at the same time. These terms are 
displayed in Fig.~\ref{fig:components} (without the multiplicative
factors involving $b$ and $f$). The BAO is only present in the
$\ell=0$ or ``real-space'' term determined by the bias. In real space the correlation becomes negative for
scales larger than $\sim 4^\circ-5^\circ$. The effect of RSD, in
particular of the cross term $b f$, is to make the correlation
positive until larger scales, broadening the BAO feature (see also Fig. 5 in \cite{2011MNRAS.tmp..385C}).
Hence, at these scales the value of $f$ can be degenerate with
``excess-power'' caused by systematic effects that also make the
correlation positive (e.g. extinction, star-galaxy separation or magnitude errors).   

\subsection{Fits to growth and bias}
\label{sec:fits}

We will now use the measurements of the angular correlation function
and the model described above to investigate the constraining power of
the SDSS LRG photometric catalog onto the two-parameter
space given by the velocity growth factor $f$ and large-scale galaxy
bias $b$. Recall however that the multipoles $\xi_\ell$ are arbitrarily
normalized to, say, the amplitude of fluctuations $\sigma_8(z=0)$ in spheres of
$8 h^{-1}\,{\rm Mpc}$ (e.g. \cite{2009JCAP...10..004S}). Thus, $f$ and $b$ are perfectly degenerate with
this normalization and our two parameter space is in fact given by $b(z) \sigma_8(z)$ and $f(z) \sigma_8(z)$, where
$\sigma_8(z) = D(z) \sigma_8(z=0)$ and $D(z)$ is the linear growth
factor.

\begin{figure}
\begin{center}
\includegraphics[width=0.4\textwidth]{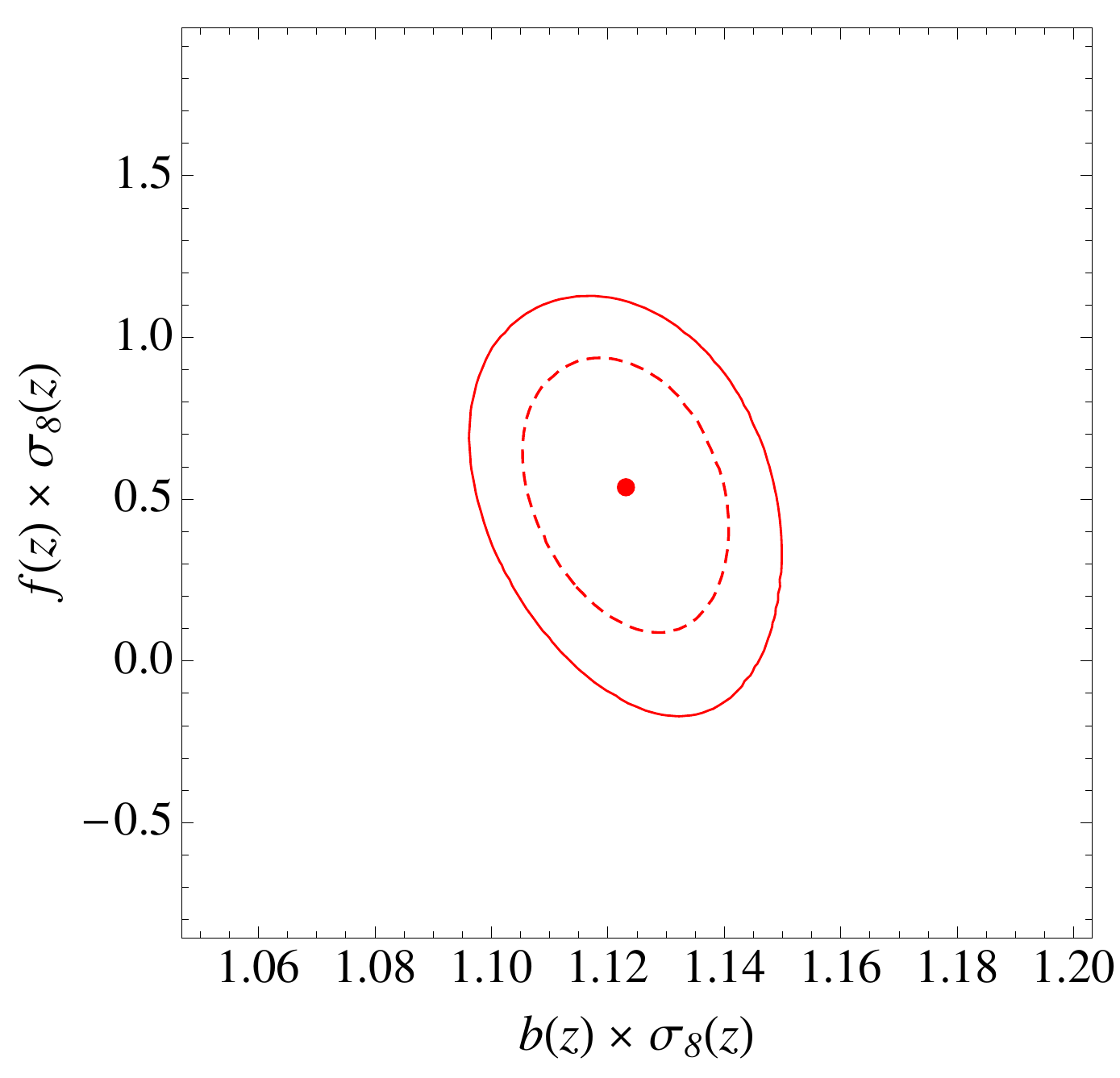} 
\caption{Two dimensional constrains in $f\times \sigma_8(z)$ and $b(z)
  \times \sigma_8(z)$. Dashed line corresponds to the $1-\sigma$ marginalized
  contour ($\Delta\chi^2=1$) and solid line to the unmarginalized case
  ($\Delta\chi^2=2.3$). In our fiducial cosmology
  $\sigma_8(z=0.5)=0.61$, leading to best-fit values $f(z=0.5)=0.87$ and $b(z=0.5)=1.84$}
\label{fig:fbcontour}
\end{center}
\end{figure}

Notice that we will not make any assumption for the relationship between the
velocity growth factor $f$ and cosmological parameters, but rather take
$f$ as a free-parameter. In particular we
will not assume that the underlying theory of gravity is General
Relativity (where to a good approximation $f = \Omega_m^\gamma(z)$ and
$\gamma=0.55$) because our ultimate goal is in fact understanding to what extent
photometric data can help to constrain GR.

This is different from other works in the literature where constraints from redshift space distortions
for $f$ (or $\beta = f/b$) are recast or combined with the ones for $\Omega_m$ assuming GR  (e.g.
\cite{2007MNRAS.378..852P}, \cite{2007MNRAS.374.1527B}, \cite{2011MNRAS.412.1669T},
see also Sec.~\ref{sec:implications}). 
Our approach corresponds to the {\it free growth} model of \cite{2011arXiv1102.1014S}.

For the cosmological model we will assume a $\Lambda$CDM universe compatible with WMAP7
(\cite{2011ApJS..192...18K}) with $\Omega_m=0.272$,
$\Omega_{DE}=0.728$, $\Omega_b=0.0456$, $n_s=0.963$, $h=0.704$
. In our approach these parameters determine the shape of the
real-space matter correlation function and the distance-redshift relation.

\begin{table} 
\begin{center}
\begin{tabular}{lccccc}
\hline \\
redshift bin   &    $b(z) \sigma_8(z)$   &   $f(z) \sigma_8(z)$        &&  ${\rm d} D/{\rm d} a$   \\  \\
$0.15-0.30$    &  $1.46 \pm 0.16$  &   $0.49 \pm 0.10$        && $0.76 \pm 0.15$  \\ 
$0.30-0.40$    &   $1.28 \pm 0.08$   &  $0.42 \pm 0.06$        && $0.70 \pm 0.10$  \\
$0.40-0.47$    &    $1.46 \pm 0.16$   &   $0.50 \pm 0.12$       &&
$0.90 \pm 0.21$  \\  \\ 
$0.50-0.60$    & $1.12 \pm 0.02$ & $0.53 \pm 0.42$ && $1.04 \pm 0.81 $  \\  \\ 
\hline
\end{tabular}
\end{center}
\caption{Best-fit values for bias, velocity growth factor and growth
  rate. Top three rows correspond to the analysis of spectroscopic
  SDSS LRG data as presented in Cabr{\'e} $\&$ Gazta{\~n}aga
  (2009). Bottom row to our single redshift bin using the photometric LRG catalog.}
\label{table:best-fit}
\end{table}

As shown in Sec.~\ref{sec:photoz} splitting the data in multiple bins results in 
samples that are highly correlated. Thus, we decided to focus in a
single redshift bin to present our main results and defer the study of narrow
bins to Sec.~\ref{sec:systematics}, in the context of robustness and consistency studies.

Hence we concentrate on the bin $[0.5-0.6]$. This
width is about twice the typical photometric error while the center of the bin makes this data uncorrelated with the
SDSS LRG spectroscopic sample (that we refer to later on). The estimate for the true redshift distribution of
galaxies in this bin is shown in Fig.~\ref{fig:dndz0506}. 
Using a spline-fit to it we computed the observed model correlation following
Eqs.~(\ref{eq:wstars},\ref{eq:wtheta}-\ref{eq:wtheta2}) sampling the
two-parameter space given by $f\sigma_8-b\sigma_8$. We then performed
a standard $\chi^2$ minimization, where
\begin{equation}
\chi^2(f \sigma_8,b \sigma_8) = \sum_{i,j=1}^{N_{bin}} \Delta w (\theta_i){\rm Cov}_{ij}^{-1} \Delta w(\theta_j)
\end{equation}
and $\Delta w \equiv w_{obs,model}(f \sigma_8,b \sigma_8) -
w_{measured}$. To begin with we use the jack-knife covariance.
The resulting best-fit values and
$1-\sigma$ errors are listed in Table~\ref{table:best-fit} (to convert
to $f$ and $b$ one can use that $\sigma_8(0.55)=0.611$ for our
fiducial cosmology if $\sigma_8(z=0)=0.8$). 
The fit yields $\chi_{min}^2= 26.67$
for $28$ bins in $\theta$ in the range
$\left[0.6^{\circ}-6^{\circ}\right]$ and two fitting parameters.
Hence the quality of the fit is very good.
The joint error distribution is
displayed in Fig.~\ref{fig:fbcontour} and shows that these two
parameters are not degenerate. The significance of RSD in our data is
$\sim 1.26 \, \sigma$ (i.e. the degree by which the recovered value
for $f$ is away from zero).

Alternatively one can assume that the velocity growth rate is given by General Relativity,
i.e. $f(z)\sigma_8(z) = \sigma_8(z){\rm d} \ln D(z) / {\rm d} \ln a = 0.45$ (for our fiducial
cosmology and $\sigma_8(z=0)=0.8$), and fit only for the overall
amplitude of clustering. 
In this case we recover a similar $\chi_{min}^2 (=26.7)$ and $b\,\sigma_8=1.125 \pm 0.017$. Hence the fit
does not vary appreciable, showing that agreement with
standard $\Lambda$CDM is also very good and that the sensitivity of our data to
redshift-distortions is weak, as expected.

\begin{figure*}
\centering
\begin{tabular}{cc}
\includegraphics[width=0.65\textwidth]{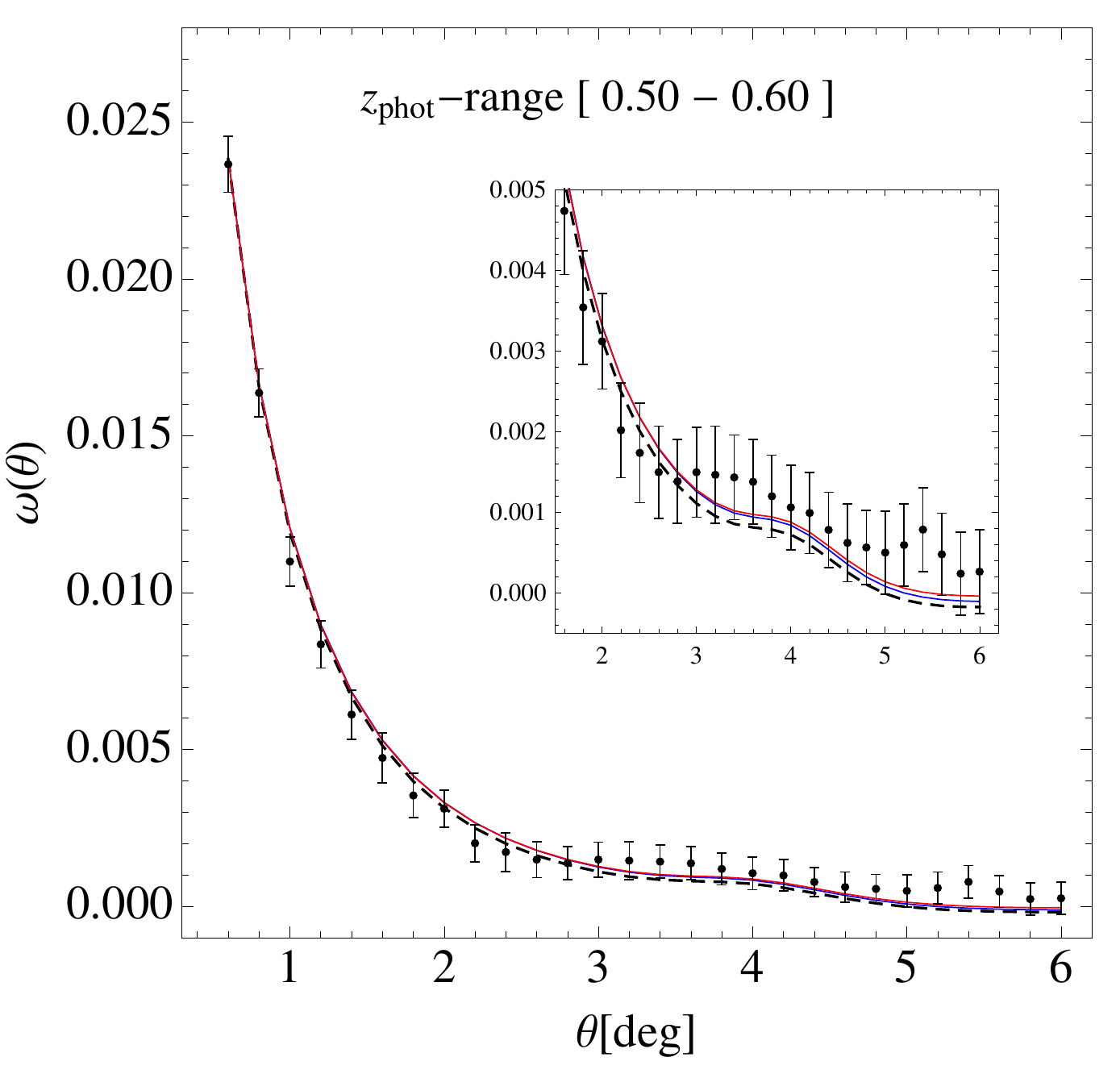} \\
\end{tabular}
\caption{Angular correlation function in our central bin of width
  $0.1$ centered at $z=0.55$. Solid red line is our best-fit model
  including the effect of star contamination ($f_{star}=4\%$). Solid blue
  line is the corresponding best-fit model if $f_{star}=0$. The values
for f and $b$ are given in Table \ref{table:best-fit}. For reference
we include with a dashed black line a WMAP7 $\Lambda$CDM model assuming General Relativity, that is
with $f$ set to $\Omega_m^{0.55}$ ($f_{star}=0$ in this case). The inset panel zooms in
the region where the baryon acoustic peak is located (see Fig.~\ref{fig:BAO}
for the BAO significance). Notably the
different models match the data very well in all the range of scales.}
\label{fig:wtheta_z050-060}
\end{figure*}

The results so far were obtained assuming $4\%$ star contamination (as
discussed in Sec.~\ref{sec:starcontamination} and Eq.~(\ref{eq:wstars})). 
If we now assume that our sample is perfectly clean of stars
($f_{stars}=0$) we get
$b \,\sigma_8 = 1.08 \pm 0.02$ and $f \,\sigma_8 = 0.66 \pm 0.39$ and
a $\chi^2_{min}=26.87$. This model is hardly differentiable from the case with
$f_{stars}=4\%$ (the $\chi^2$ becomes only negligible worse), see Fig.~\ref{fig:wtheta_z050-060}.
Hence stellar contamination can mimick the effect of RSD and become a
worrisome source of systematic effect in future and better data.
In the present case the agreement between the best-fit value of $f$ and the ``GR'' value
degrades only slightly when $f_{star}=0$, with differences well within
the errors. We conclude that the star contamination in our sample is
sufficiently under control and does not drive our results.

Figure \ref{fig:wtheta_z050-060} shows the resulting best-fit models
discussed above against the measured angular correlation function. 
In all cases the model matches the data quite well
in all the angular range, with no signal of excess-clustering on large
scales. This is to some extent at variance with the recent study by
\cite{2011MNRAS.412.1669T} (see also the follow up
\cite{2010arXiv1012.2272T}) who finds the amplitude of the angular
power spectra of photometric LRGs to have a 2-$\sigma$ excess
clustering away from the $\Lambda$CDM prediction at the lowest multipoles.
This result was obtained using the MegaZ catalog over three
redshift bins in the range [0.4-0.65]. The MegaZ catalog is also build upon
the DR7 SDSS photometric catalog. However the LRG selection criteria in
MegaZ is different from the one in this paper, most notably by the
magnitude cuts ($i_{\rm deV} < 19.8$ in MegaZ compared to $petror < 21$ in our case)
\footnote{The MegaZ selection is done to match the magnitude cut in 2SLAQ from where the
photo-z calibration is obtained. We find that the angular correlation
of galaxies in MegaZ matches the clustering amplitude in
our sample if we cut in $petror < 20.7$. Hence MegaZ is slightly brighter
than our sample.} but also in the color and $mag_{50}$ cuts intended to isolate stars from galaxies.
The photo-z code used for MegaZ is also different from ours, although
both are based in the ANNz code of \cite{2004PASP..116..345C}.
\cite{2009arXiv0912.0511S} also finds an excess clustering when
searching for the baryon acoustic scale, but this result is 
difficult to compare  with ours as they use an over simplified
method to stack the signal from wider redshift bins calibrated with different spectroscopic samples.

The constraints recovered so far, albeit large, match the
study of \cite{2011arXiv1102.0968R} that forecast how well redshift
distortions can potentially be measured with galaxies selected from a photometric survey.
This match can be clearly seen from their Fig. 8 after scaling the
results for a sample of unbiased galaxies binned around $z=0.5$ by our
bias factor $b\sim2$ (since $\Delta(f\sigma_8) \propto b$). For our
redshift bin width the forecast yields $\Delta(f\sigma_8) \sim
0.4-0.5$, in very good agreement with the results we obtain with the
actual data (e.g. see Table \ref{table:best-fit}). 
This implies that systematics, photo-z errors, selection and modeling can be
controlled sufficiently well in present photometric data to 
yield expected constraints of redshift distortions. Near future data,
provided with a better handle on systematics due to stars and photo-z, 
should be able to complement shallower spectroscopic surveys in placing bounds to the growth of
structure in the universe.

\begin{figure}
\begin{center}
\includegraphics[width=0.4\textwidth]{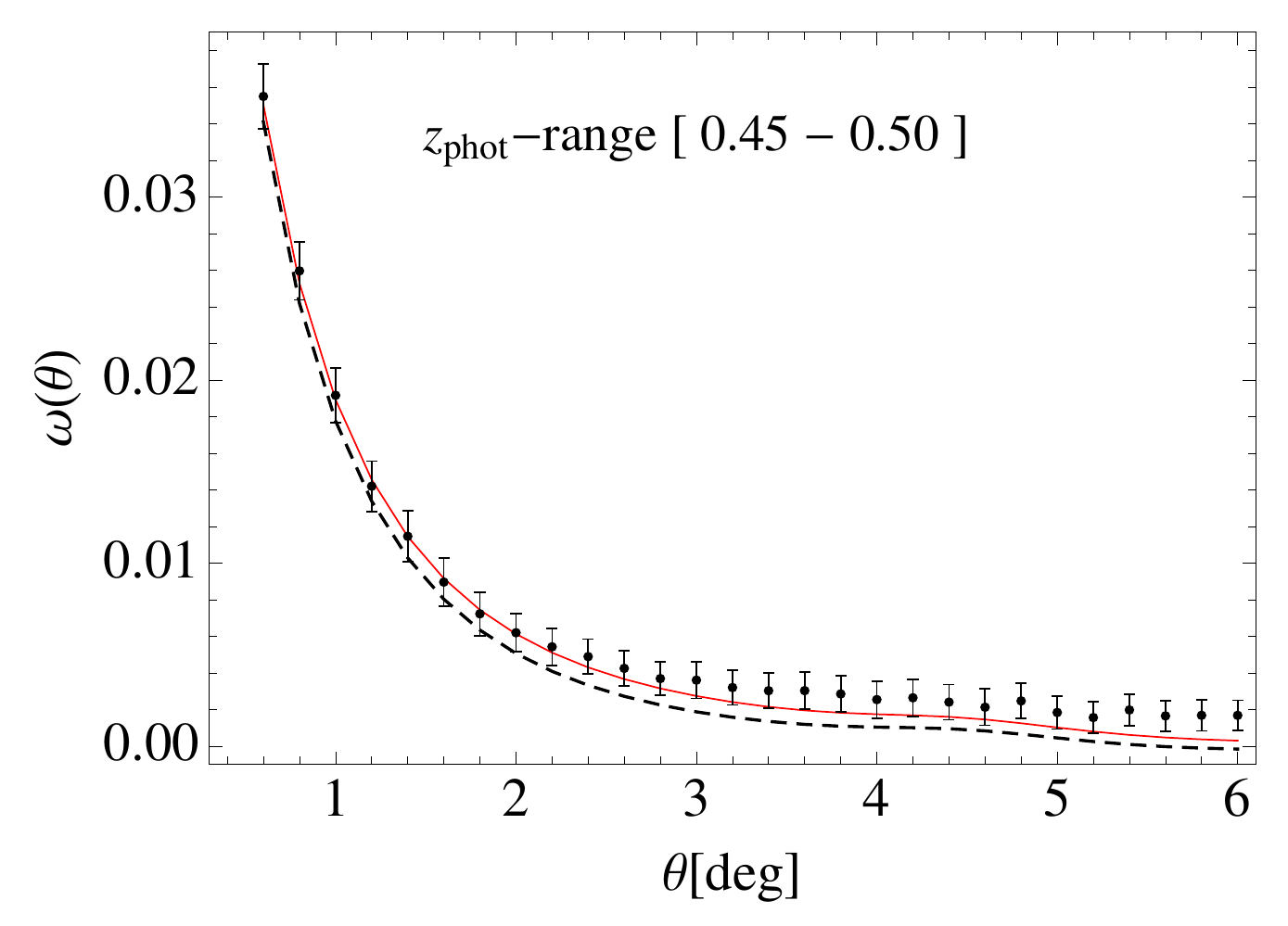} \\
\includegraphics[width=0.4\textwidth]{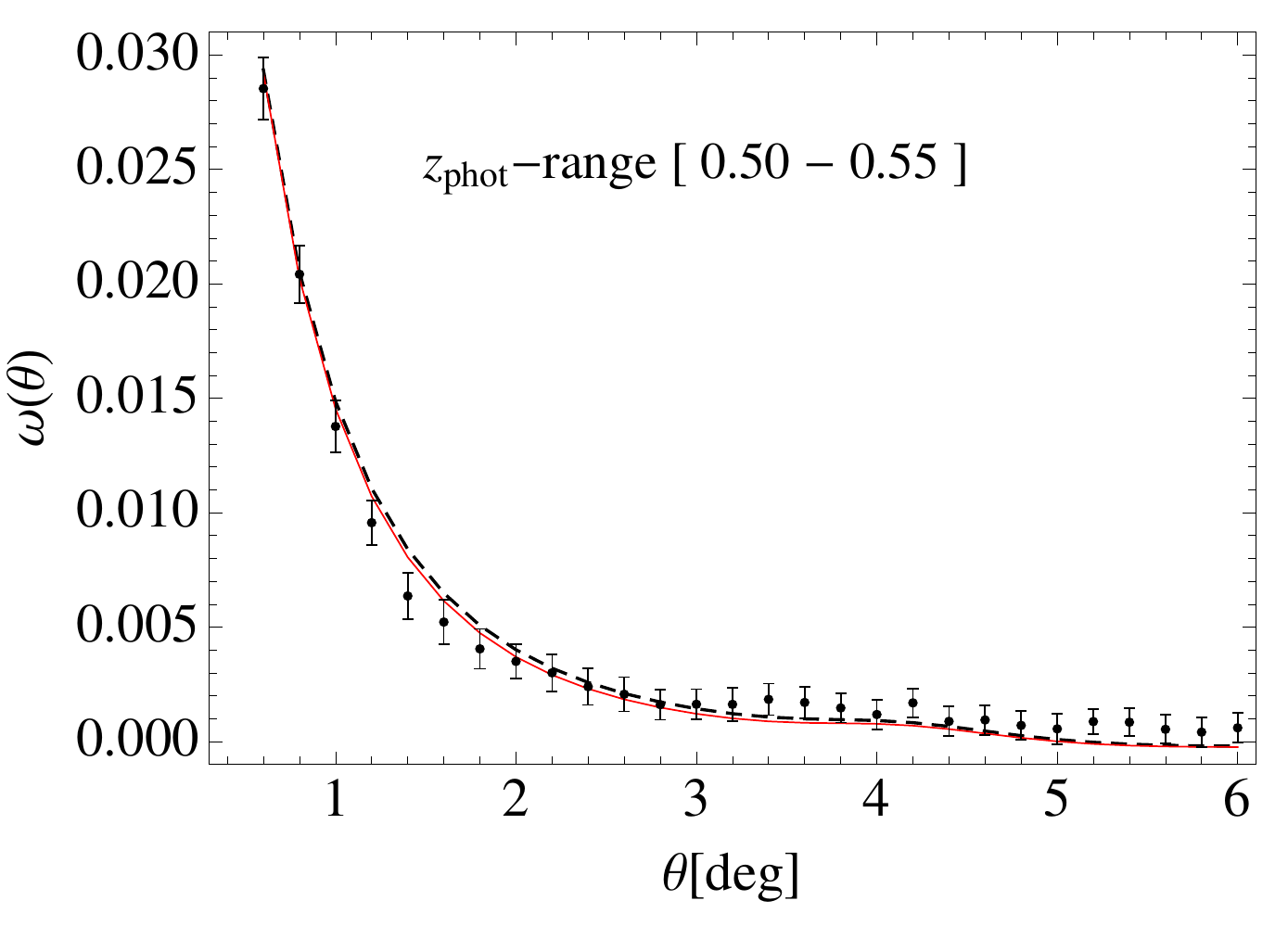}  \\
\includegraphics[width=0.4\textwidth]{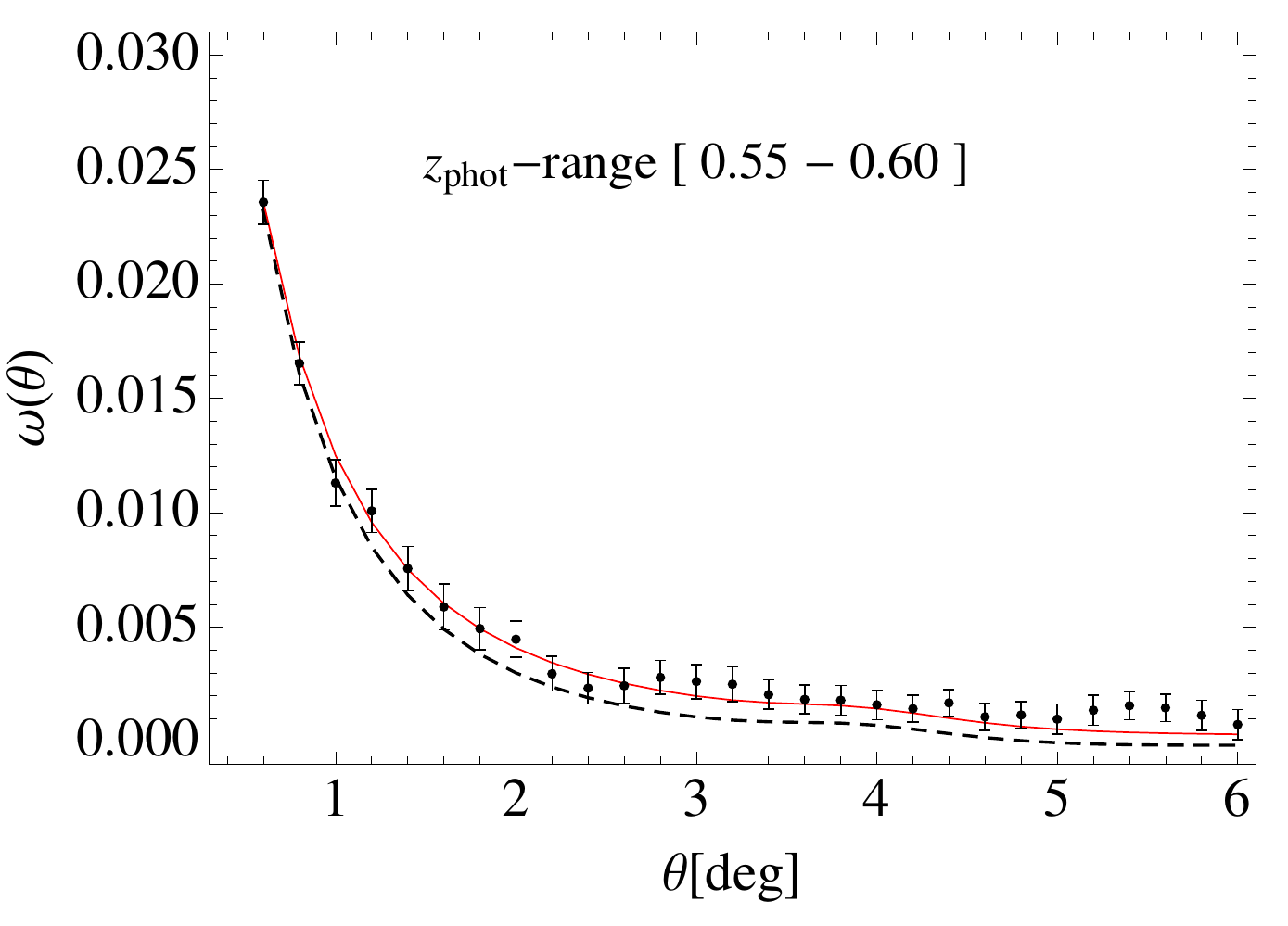} 
\caption{Angular correlation function in ``narrow'' bins of width
  comparable to the underlying photometric error, as indicated
  in the panels. Solid red lines are the best-fit models taking into
  account the $\sim 4\%$ star contamination in our sample. Dashed line correspond to a standard
$\Lambda$CDM model with $f_{star}=0$ and growth given by General
Relativity, i.e. $f(z)=\Omega_m(z)^{0.55}$.}
\label{fig:wtheta_narrowbins}
\end{center}
\end{figure}

\subsection{Robustness and impact of systematics effects}
\label{sec:systematics}

In this section we investigate different components of our analysis
that could potentially change our results.
\newline

{\it Analytical Covariance Matrix:}
We now investigate how our results change when we use the analytical
error estimate discussed in Sec.~(\ref{sec:errors}). This estimate accounts for statistical and shot-noise
errors but does not account for 
systematic errors introduced by, say, stars (at least in the way
presented in \cite{2011MNRAS.tmp..385C}). Hence we compute the $C_{\ell}$
spectra, and ${\rm Cov}_{Th}$, in Eq.~(\ref{eq:covT}) using the best-fit values for $f$ and $b$
corresponding to the model with $f_{stars}=0$ discussed above.
Using this covariance the best-fit values to the angular correlation
in the bin $\left[0.5-0.6\right]$ change to $b \sigma_8 = 1.13 \pm 0.02$ and
$f \sigma_8 = 0.35 \pm 0.54$ while the $\chi^2_{min}$ degrades to
$32$ (for a model with $f_{stars}=4\%$). 
Hence the change in the
recovered best-fit values is within half $\sigma$ compared to the results shown
in Table \ref{table:best-fit}. In turn, the error in $b
\sigma_8$ is unchanged while that in $f \sigma_8$ increases
by about $25\%$.

The resulting model matches the data as well as the one derived using
jack-knife errors and the goodness of the fit are comparable. The only
difference is in the resulting error on the velocity growth factor. 
This may be due to the structure of the covariance on large
separations that may be affected by some
systematics not captured by the theoretical estimate.
\newline

{\it Redshift Distribution:}
One important but difficult to estimate component in clustering analysis of photometric
data is the distribution of galaxies in true redshift. In
Sec.~\ref{sec:photoz} we studied the degree of uncertainty left in the
estimate of $N(z)$, about $1\%$ in the peak position
and $9\%$ in the width. To investigate the impact of this systematic
in our analysis we computed the model correlation and found best-fit
parameters assuming $N(z)$ is a
Gaussian distribution with $\mu=0.549$ and $\sigma_z=0.062$ (model-1),
that are the
best-fit values to $N(z)$ in Fig.~\ref{fig:dndz0506}. We then
recomputed the model by increasing
$\mu$ by $1\%$ and decreasing $\sigma_z$ by $9\%$ (model-2), and
subsequently did the best-fit analysis. The $\chi^2$ of
these two models is almost equal, changing by $\sim 1\%$. The
resulting best-fit values for $b\sigma_8$ decrease from $1.12 \pm
0.02$ (for model-1)
to $1.08 \pm 0.02$ (for model-2) while from $0.76 \pm 0.57$ to $ 0.7\pm 0.5$ for $f\sigma_8$. This is
because a narrower $N(z)$ implies less bin proyection what leads to a 
higher amplitude angular correlation and slightly more
sensitivity to redshift distortions (\cite{2010MNRAS.407..520N,2011MNRAS.tmp..385C}).
These differences are only marginal given the overall large errors provided by present photometric data.
However future surveys  will probably need more accurate estimates of $N(z)$.
\newline

{\it Narrower redshift bins:}
As mentioned earlier we have also considered splitting the data into 3
narrower bins of width similar to the typical photo-z error : $\left[0.45-0.5\right]$,
$\left[0.50-0.55\right]$ and $\left[0.55-0.60\right]$.
The galaxy redshift distributions for these bins are shown in
Fig.~\ref{fig:dndz_narrowbins} while the measured correlations and
best-fit models are displayed in Fig.~\ref{fig:wtheta_narrowbins}. 

The best-fit values for the bias in each bin decreases slightly with
increasing redshift: $b \sigma_8 = 1.26,
1.21$ and $1.1$ respectively (with $2\%$ error).

In turn the best-fit values for $f \sigma_8$ are $1.14\pm 0.57$, 
$0.024\pm 0.53$ and $1.39\pm 0.46$ respectively (assuming $f_{star}=4\%$).
Hence we see some spread in the recovered values for the
velocity growh rate. Compared to the corresponding values in GR
they still agree at $\sim 1-2\,\sigma$. 

As discussed before a bin width smaller or comparable to the photo-z is subject to large
bin to bin migration (in other words, the estimate of $N(z)$ itself is
more sensitive to photo-z unkowns). Therefore we expect to recover more robust
results in bins larger than the underlying photo-z.
\newline

\subsection{Implications for the growth rate and comparison with
  spectroscopic studies}
\label{sec:implications}

We now put our results in context of similar ones derived using spectroscopic data.
In Table \ref{table:best-fit} we show constraints in $b(z) \sigma_8(z)$ and
$f(z)  D(z)$ as obtained by \cite{2009MNRAS.393.1183C} using
spectroscopic LRGs from SDSS.
Table 2 in  \cite{2009MNRAS.393.1183C} list the best fit values of the clustering amplitude
$Amp\equiv b(z) \sigma_8$ and $\Omega_m$. This Table refers 
to $\sigma_8$ at $z=0$ because the best fit value of $\Omega_m$ was used to
estimate the linear growth $D(z)$ at the corresponding redshift according
to standard cosmological equations in General Relativity (GR). Here we do not
want to assume GR or any other relation between $\Omega_m$ and $D(z)$
and we therefore scale these amplitudes back to the original data
amplitudes, i.e.  $\sigma_8(z)$, by multiplying $Amp$ by the best fit
value of $D(z)$ in GR. The resulting amplitudes are listed here as $b(z) \sigma_8(z)$ in 
Table 1. We can then find an estimate of $f(z) \sigma_8(z)$ by
just multiplying these $b(z) \sigma_8(z)$  estimates with the values of $\beta$ listed
in Table 1 of \cite{2009MNRAS.393.1183C}  for the corresponding samples.

\begin{figure}
\begin{center}
\includegraphics[width=0.45\textwidth]{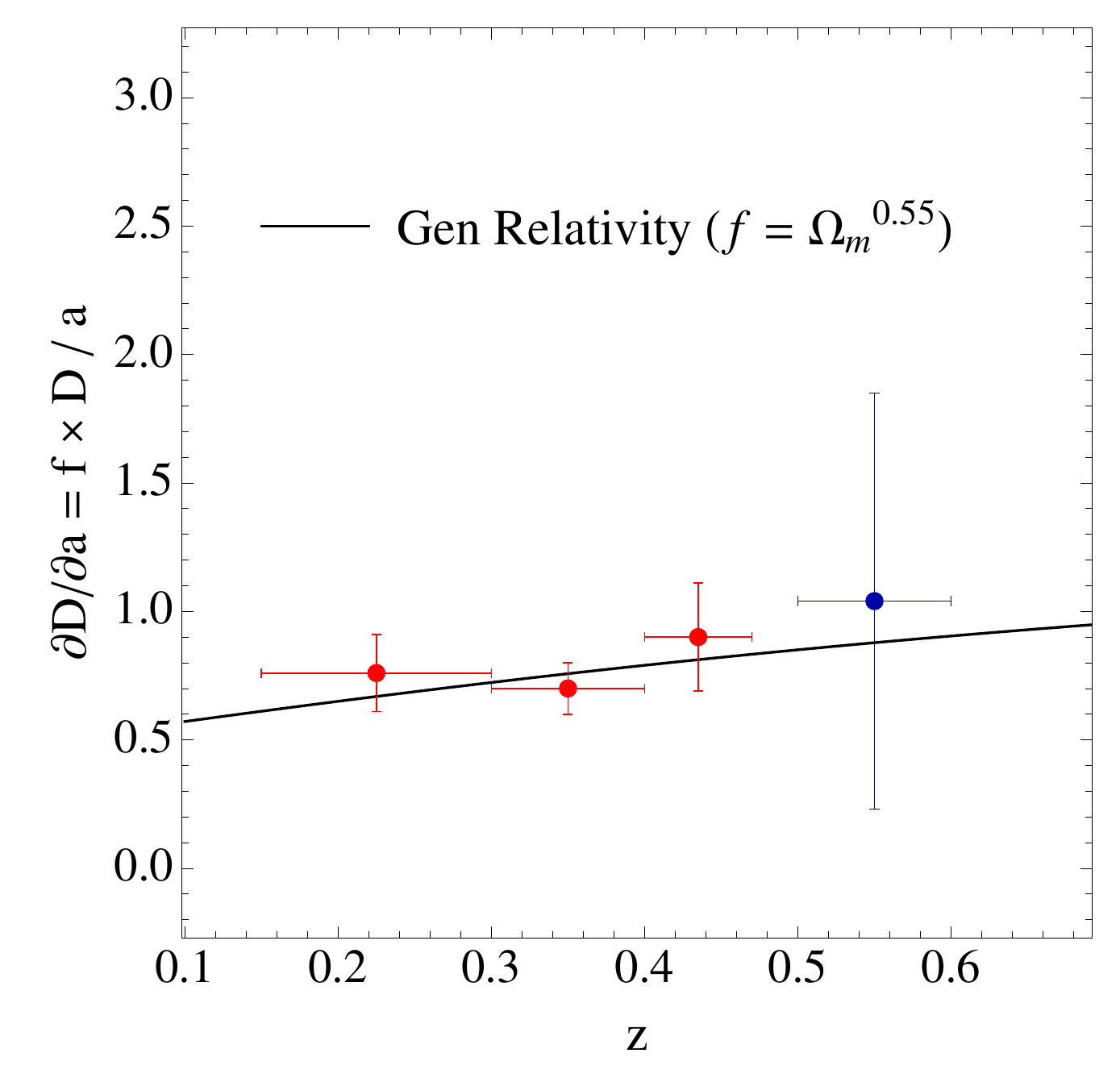} 
\caption{Linear growth rate of structure from LRG spectroscopic data in
  the range [0.15-0.47], as
  presented in Cabr{\'e} $\&$ Gazta{\~n}aga
  (2009), and from our analysis of photometric data at $z=0.55$
  (assuming $\sigma_8(0)=0.8$).
These data leads to $\gamma = 0.54 \pm 0.17$ in a model where $f=\Omega_m(z)^\gamma$.}
\label{fig:dDda}
\end{center}
\end{figure}

One can turn these values, and those at $z=0.55$ from the photometric data, into estimates for the linear growth rate of structure
as follows,
\begin{equation}
\frac{\partial D}{\partial a} = \frac{D(z)}{a} f(z) =
\frac{1+z}{\sigma_8(0)} f(z)\sigma_8(z),
\end{equation}
where we assume our fiducial value $\sigma_8(0)= 0.8$ (consistent
with, e.g., \cite{2011arXiv1104.1635T}).
Results are listed in Table~\ref{table:best-fit} and displayed altogether in Fig.~\ref{fig:dDda}.
Provided with them one give constraints in the growth index assuming the $\gamma$-parameterization of the growth,
for which $f=\Omega_m(z)^\gamma$ and $\gamma$ is a scale and redshift independent constant, \cite{2005PhRvD..72d3529L}.
In this case,
\begin{equation}
\frac{\partial D}{\partial a} = (1+z) D(z) \Omega_m(z)^\gamma
\end{equation}
and $D(z)=\exp{(-\int_0^z \Omega_m(x)^\gamma/(1+x) dx)}$. This yields
\begin{equation}
\gamma = 0.54 \pm 0.17
\end{equation}
for the combination of spectroscopic and photometric data given in the
fourth column of Table \ref{table:best-fit}.

Before moving on we note that novel constraints on the
growth rate up to $z=0.9$ were very recently reported by the WiggleZ survey using blue galaxies
instead of LRGs (\cite{2011MNRAS.tmp..834B}). At $z<0.5$ they
 improve to some extent over the ones we used, e.g. $\Delta(f \sigma_8)=0.07$
 at $z=0.2$ and $\Delta(f \sigma_8)= 0.04$ at $z=0.4$ (see
 Table~\ref{table:best-fit}). At $z>0.5$ they are considerably tighter
 than those we derive with imagining data, as expected.
 Yet, for concreteness we have decided to present our results in terms of SDSS LRG clustering
 (either spectroscopic or photometric) as the focus of the paper is to
 demonstrate that imaging data is able to yield constraints in RSD.

\section{Baryon Acoustic Oscillations}
\label{sec:bao}

One of the most important probes of the accelerated cosmic expansion is the 
existence of an excess clustering imprinted at a well defined
comoving length scale of about
$\sim 100\,{\it h}^{-1}{\rm Mpc}$ in the correlation of galaxies and 
$\sim 1^\circ$ in the one of CMB photons. It originates in the coupling
of the baryon-photon plasma prior to recombination and hence can be 
very well determined with CMB data. Provided with this estimate it can then be
measured with galaxy data and used to constrain the distance-redshift
relation in the local universe, that in turn is sensitive to the
nature of dark-energy (and/or the appropriate law of gravity).
The main obstacle of this pathway is that the excess clustering signal
in galaxies represents only $\sim 1\%$ over that of a random distribution.
Nonetheless several surveys are dedicated or include BAO in
their scientific plans.

The BAO signature have been extensively studied in the clustering of
spectroscopic data, in particular using LRGs since they span the
largest volume possible compared to other galaxy types
(e.g. \cite{2005ApJ...633..560E,2006A&A...449..891H,2007MNRAS.381.1053P,2009MNRAS.399..801Gazta,2009MNRAS.393.1183C,
2009MNRAS.399.1663G,2010ApJ...710.1444K,2010MNRAS.401.2148P}). 
The advantage in this case is that one has three
dimensional information to sample this $1\%$ excess of pairs. 
The disadvantage is that the sampling of the over-density
field is much more poorer with spectra and is eventually limited to
lower redshifts when compared to photometric catalogs (which on the
converse only yield projected quantities with lower significance). 
Hence it is very important to investigate what evidence for this
signature is already present in our data given the number of such
photometric surveys already undergoing or planned for the near-future. 

The use of photometric data to investigate BAO has been relegated to some extent, probably due to the lower
signal-to-noise, the impact of systematics and the quality of the photo-z.
In \cite{2007MNRAS.374.1527B} and \cite{2007MNRAS.378..852P} this signature was studied stacking photometric redshift
bins in order to reconstruct the 3D spectrum. Both studies find
evidence for BAO at $< 3\,\sigma$. In turn, only \cite{2009arXiv0912.0511S} (to
our knowledge) have explored the possibility of locating the signature
in configuration space, with ambiguous results (e.g. too high in
amplitude compared to $\Lambda$CDM expectations).

We will now investigate how well our measurements agree with the shape
of the model correlation including or excluding the effect of
baryons. In a separate paper we present a detailed analysis of 
the significance and implication of the baryon acoustic peak imprinted
in the angular clustering of our sample, see \cite{bao_paper}.

\begin{figure}
\begin{center}
\includegraphics[width=0.47\textwidth]{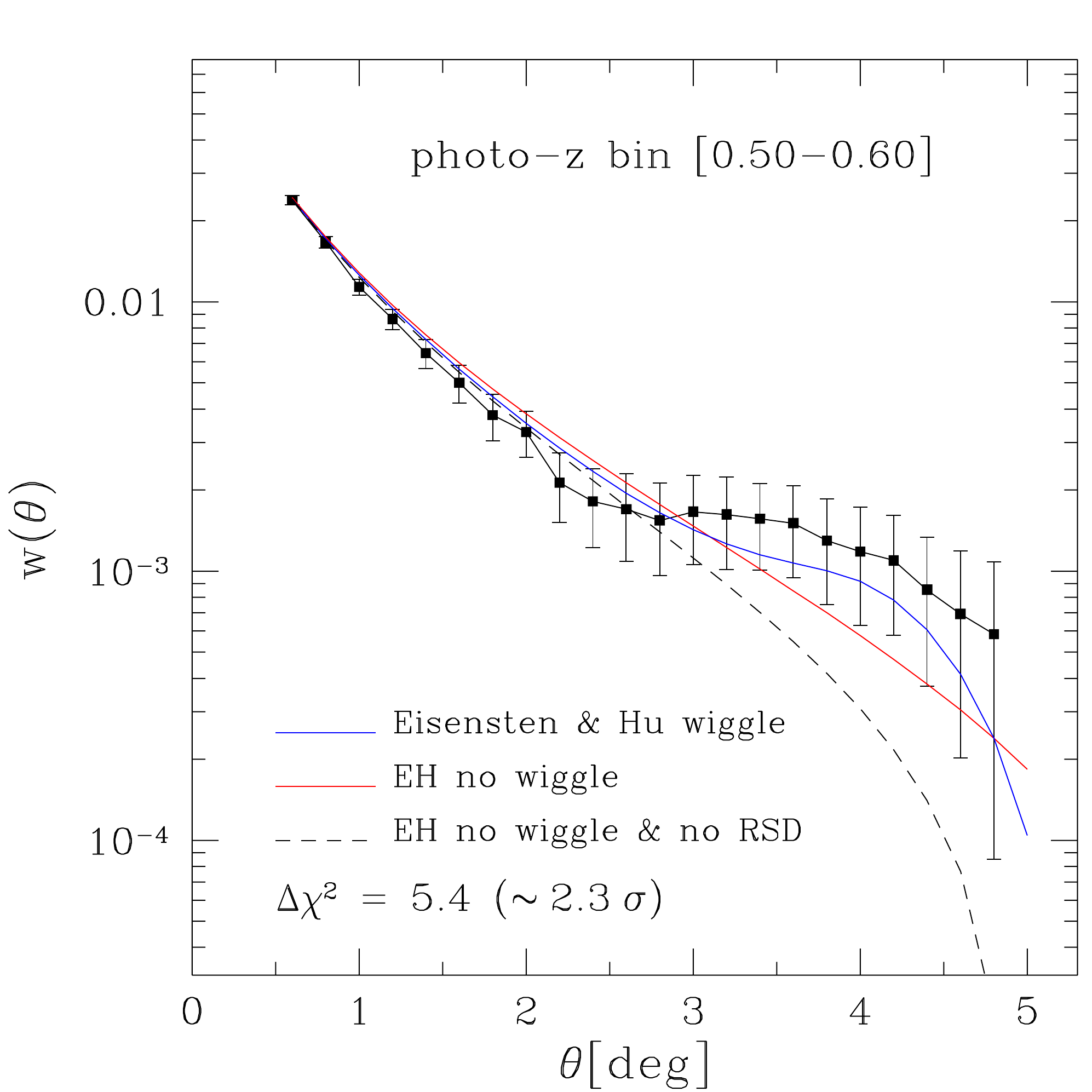} 
\caption{{\it Baryon Acoustic Feature at $z=0.55$}. The plot shows
  the Eisenstein $\&$ Hu (1998) model with and without baryons. The
  data and model agree quite well in both amplitude and shape of the 
  acoustic bump. The BAO feature is detected here with a $98\%$
  confidence level ($2.3 \,\sigma$ significance). Dashed line corresponds to the no-wiggle model when RSD are
  ignored. This shows the importance of both BAO and RSD to account
  for observations.}
\label{fig:BAO}
\end{center}
\end{figure}

Figure~\ref{fig:BAO} shows the measured angular correlation in the
photo-z bin $\left[0.5-0.6\right]$ in a way that highlights the 
shape of $w(\theta)$ at large angular scales. We also display the
standard \cite{1998ApJ...496..605E} model prediction including the baryons
effect (solid blue) and excluding it, i.e. with the wiggles smoothed
out (solid red). Figure~\ref{fig:BAO} shows a quite clear bump in the 
correlation function at $\theta \sim 4^\circ$, in very good agreement 
with the model in both shape and location. For this figure we have not
attempted to fit for redshift distortion parameter $f(z)$ as in Sec.~\ref{sec:fits} but rather have
assumed GR ($f(z=0.55)=0.74$) and fit for the overall bias only
(with fixed cosmological parameters as given in Sec.~\ref{sec:fits}). We note that the ``BAO'' model correlation obtained with 
the approximation of \cite{1998ApJ...496..605E} is very similar the
one derived with CAMB in Sec.~\ref{sec:fits}.

To estimate the statistical significance of this feature we compare 
the two best-fit models, with and without BAO. The difference in their $\chi^2$ yields $5.4$ 
suggesting that the BAO model is preferred with a significance $\sim
2.3 \,\sigma$ ($\sim 98\%$ confidence). Notice that both models have the same number of degrees
of freedom since only the bias is fit. This result is un-changed if we
allow for $4\%$ star contamination in both models and is in perfect
agreement with the finding of \cite{bao_paper} using a completely
independent analysis method. 

Although the significance is low the good agreement with $\Lambda$CDM
expectations is very encouraging for future photometric campaign that will achieve better  
error bars thanks to improved photo-z and survey depth. In addition
this is, to our knowledge, the first time the BAO bump is clearly
depicted in agreement with $\Lambda$CDM predictions in the angular correlation of photometric LRG data at this
high redshift.

\section{Conclusions}
\label{sec:conclusions}

In this paper we used the angular correlation function of the imaging
sample of luminous red galaxies in the DR7 of the SDSS as a testing
ground for measuring the growth rate of expansion 
through redshift space distortions using photometric data. In addition
we investigated the evidence for the baryon acoustic feature in the
angular correlation.

We put a strong emphasis in the selection of the galaxy sample
and overall robustness against several systematic effects, such as
magnitude errors, bad extinction zones, photo-z outliers and more.
We paid particular attention to the impact of stellar contamination
in the angular clustering measurements. On the one hand we
minimized such contamination introducing a cut in surface brightness,
$mag_{50}$. On the other we estimated the distortion that 
the residual contamination introduces in the correlation measurements.

Our measured correlation, in the range $\left[0.50-0.60\right]$ is in
good agreement with expectations from standard $\Lambda$CDM. In
particular it shows no excess clustering on the largest scales, contrary to other works in the
literature. This is a encouraging proof that systematic effects in 
photometric data can be controlled sufficiently well to use them as a
cosmological tool
(\cite{2007MNRAS.374.1527B,2007MNRAS.378..852P,2011MNRAS.412.1669T}). 
The distortions introduced by the intrinsic
clustering of contaminating stars does not change our results but 
might become a worrisome source of systematic biases in future
surveys with smaller error bars. A similar conclusion was reached 
in regards to the estimation of the true redshift distribution of the sample.

Indeed, we found that redshift space distortions can be measured using
photometric data, albeit with large error bars due to the high-bias of
the sample and the poor photo-z error. Our results are in very good
agreement with the recent forecast by \cite{2011arXiv1102.0968R}. Hence, this
paper can be taken as a validation of the forecast for a SDSS like-case
 as presented in \cite{2011arXiv1102.0968R} and a proof-of-concept of
 the promising expectations for upcoming photo-z surveys such as DES, Euclid and PannStars.

In addition we found quite a strong evidence for the baryon acoustic peak in the
measured angular correlation, something not observed before. 
The shape, amplitude and location of the BAO feature is in very good
agreement with $\Lambda$CDM expectations yielding a $\sim 2.3\,\sigma$
significance over a model without BAO. In a separate work, \cite{bao_paper},
we discuss this detection of BAO and its
cosmological implication using an independent analysis from the one
presented here.

In all, our results strengthen the expectations on the ability of
future photometric surveys to compete and/or complement spectroscopic
data, as well as to serve to other approaches such us weak lensing or
supernovae, in the quest to understand the nature of cosmic acceleration.

\section*{Acknowledgements}
\label{sec:acknowledgements}

We are thankful to Carlos Cunha for his help with photo-z.
Funding for the SDSS and SDSS-II has been provided by the 
Alfred P. Sloan Foundation, the Participating Institutions, the 
National Science Foundation, the U.S. Department of Energy, the 
National Aeronautics and Space Administration, the Japanese 
Monbukagakusho, the Max Planck Society, and the Higher Education 
Funding Council for England. The SDSS Web Site is 
http://www.sdss.org/.

    The SDSS is managed by the Astrophysical Research 
Consortium for the Participating Institutions. The Participating 
Institutions are the American Museum of Natural History, Astrophysical 
Institute Potsdam, University of Basel, University of Cambridge, Case 
Western Reserve University, University of Chicago, Drexel 
University, Fermilab, the Institute for Advanced Study, the Japan 
Participation Group, Johns Hopkins University, the Joint Institute 
for Nuclear Astrophysics, the Kavli Institute for Particle 
Astrophysics and Cosmology, the Korean Scientist Group, the Chinese 
Academy of Sciences (LAMOST), Los Alamos National Laboratory, the 
Max-Planck-Institute for Astronomy (MPIA), the Max-Planck-Institute 
for Astrophysics (MPA), New Mexico State University, Ohio State 
University, University of Pittsburgh, University of 
Portsmouth, Princeton University, the United States Naval 
Observatory, and the University of Washington.
 
We thank the Spanish Ministry of Science and Innovation (MICINN) for
funding support through grants AYA2009-13936-C06-01, 
AYA2009-13936-C06-03, AYA2009-13936-C06-04 and through the Consolider 
Ingenio-2010 program, under project CSD2007-00060. 

\bibliography{rsd_sdss_dr7}

\appendix
\section{Excess power for $z>0.6$}
\label{sec:appendix}

One frequent issue when analyzing the clustering of
LRGs is the fact that at the largest scales (e.g. BAO) the amplitude of clustering appears generally high when
compared to standard $\Lambda$CDM models
(\cite{2005ApJ...633..560E,2007MNRAS.374.1527B,2007MNRAS.378..852P,2010arXiv1012.2272T,2011MNRAS.412.1669T,2009arXiv0912.0511S,2008ApJ...676..889O,2010ApJ...710.1444K,2009MNRAS.393.1183C,2009MNRAS.400.1643S,2011arXiv1102.1014S}). The significance of this
mismatch is however uncertain since these scales are expected to be
the most sensitive ones to different systematic uncertainties as well
as the ones with largest statistical variance. 

Within the context of photometric data as in our work, \cite{2011MNRAS.412.1669T,2010arXiv1012.2272T} recently found $\sim 2 \sigma$ excess in the lowest
multipoles of the angular power spectrum of LRGs in the
MegaZ catalog of DR7 SDSS. As discussed in Sec.~\ref{sec:results} we do not find such a
discrepancy in our catalog. More so if one recalls that in configuration space data points are
expected to co-variate to some level. Potentially much more worrisome
is their finding of $\sim 4\sigma$ excess for the bin
$\left[0.6-0.65\right]$ (see also \cite{2007MNRAS.374.1527B}). 
\cite{2010arXiv1012.2272T} performed a series of checks for systematic errors but none was
conclusively the source of such an effect what led them to 
speculate with the possibility that this could be due to the imprint
of ``new physics'', such as primordial non-Gaussianities,
modifications of gravity or clustering dark energy.

\begin{figure}
\begin{center}
\includegraphics[width=0.47\textwidth]{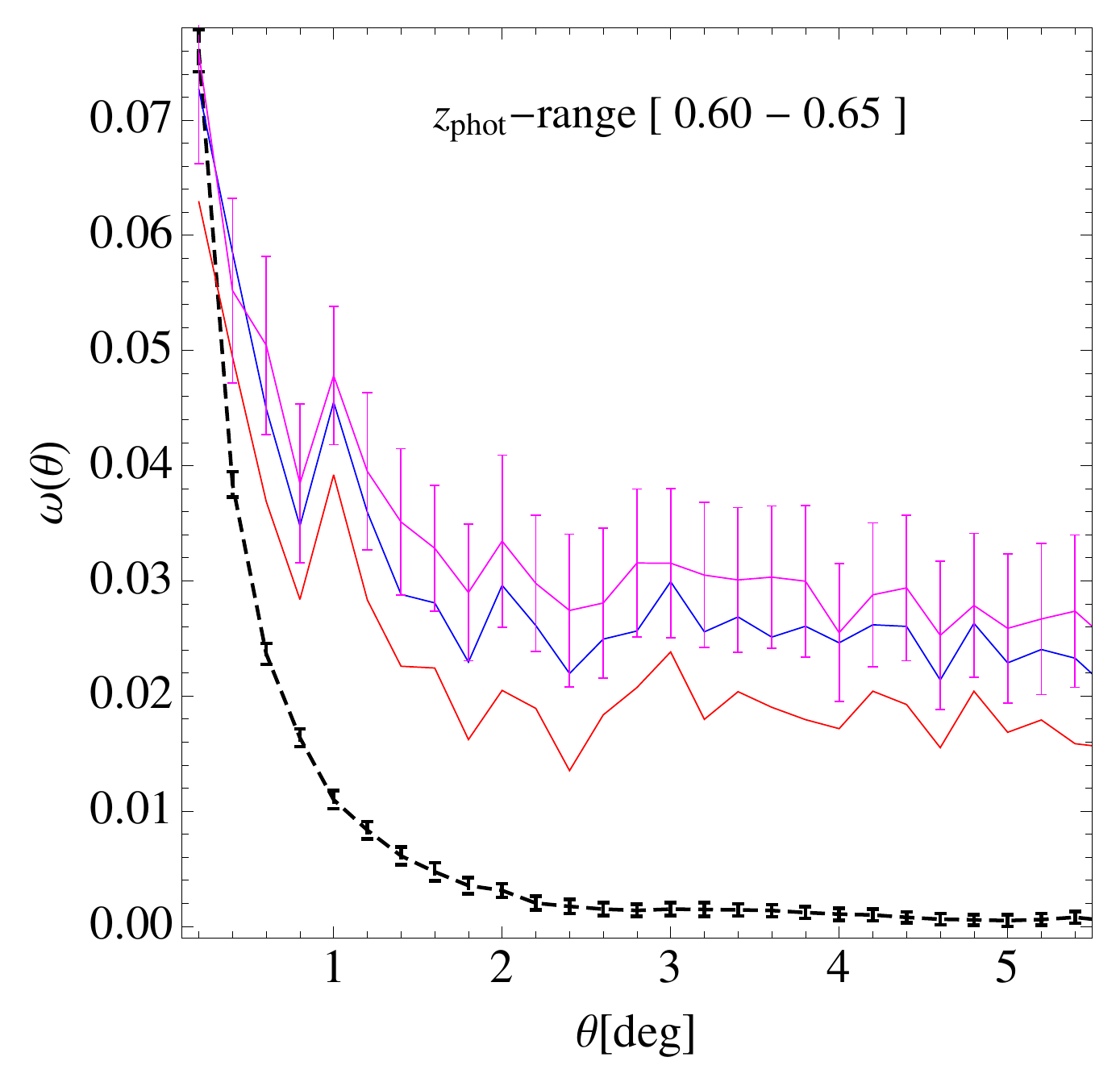} 
\caption{Excess power beyond $z=0.6$ and sensitivity to systematic
  effects. Top magenta line corresponds to a mask with no cut in
  galactic latitude or photo-z quality. Middle blue line correspond to
a cut in photo-z quality as discussed in Sec.~\ref{sec:photoz}. Red solid line
corresponds, in addition, to a mask excluding galactic latitude below
$25^\circ$, see Sec.~\ref{sec:sel}. For reference we include the measured correlation in the
bin [0.5-0.6]. The correlation seems anomalous showing an excess power 
on all scales and a large sensitivity to various systematic effects.}
\label{fig:excesspower}
\end{center}
\end{figure}

In Fourier space this excess may not be prejudicial
because it only affect few low-$\ell$ multipoles that can be cut-out
of the analysis (they have the lowest signal-to-noise anyway). When translated to
configuration-space this impacts a broad range of scales. Hence
it is interesting to see how does the clustering in configuration
space looks like at $z \ge 0.6$ to complement the study of \cite{2011MNRAS.412.1669T,2010arXiv1012.2272T}.

In Fig.~\ref{fig:excesspower} we show the measured angular correlation function for
our photometric bin $\left[0.6.-0.65\right]$. Evidently the clustering
signal does not only show an excess power at BAO scales but is in
fact anomalous at all scales (except perhaps $\theta <
0.5^\circ$). On the one hand the number of objects in this bin is
small ($\sim 30,000$) and dominates the error budget (shot-noise). On the other hand the
correlation is very sensitive to various systematic
uncertainties. This is reflected in Fig.~\ref{fig:excesspower} :
the solid magenta line corresponds to the measured correlation
when no cut in galactic latitude of the sample or photo-z quality is
imposed. Blue solid line is the result when objects with ``bad'' photo-z
are discarded from the sample (as discussed in Sec.~\ref{sec:photoz}). Lastly,
solid red line is the result when the mask is reduced by leaving out low galactic
latitudes $b < 25^\circ$ (to avoid star contamination, see Sec.~\ref{sec:sel}). One could continue with
more stringent constrains but the signal does not approach the one at
lowest bins (shown with black symbols). Displayed error bars in this
figure were obtained with jack-knife estimate.

This result signals to unknown systematic uncertainties as the most probable
cause for the anomalous shape. One possibility could be an
incomplete treatment of the distortions introduced by star
contamination. This is generally focused in reducing the fraction of
stars to the minimum possible but not to account of the residual
clustering signal, as in Sec.~\ref{sec:starcontamination}. The
contamination by stars results in a change of the mean density across the
survey area (and towards the galactic plane). This would impact only the low-$\ell$
spectra that encodes the mean density information but a broad range
of scales in the angular correlation (as shown in
Fig.~\ref{fig:wtheta_stars}). However testing this in Fourier space is beyond the
scope of this paper and will be discussed elsewhere.

\end{document}